\documentclass[a4paper, 11pt]{amsart}

\usepackage{amsmath, amssymb, amsthm}

\usepackage[boxed]{algorithm}
\usepackage{algorithmic}
\usepackage{verbatim}

\usepackage{fullpage}
\usepackage{color}
\usepackage{array}
\usepackage{pstricks}

\usepackage{natbib}

\usepackage{epsfig,subfigure}
\usepackage[normalem]{ulem}
\graphicspath{{figs/}}

\allowdisplaybreaks[4]

\newtheoremstyle{example}{\topsep}{\topsep}%
     {}
     {}
     {\bfseries}
     {.}
     {  }
     {}

\newtheorem*{thm*}{Theorem}

\newtheorem*{def*}{Definition}
\theoremstyle{example}

\newcommand{\deff }{\,\buildrel \text{\tiny def}\over{=}\,}

\newcommand{\R}{\mathbb{R}}
\newcommand{\Q}{\mathbb{Q}}

\renewcommand{\P}{\mathbb{P}}
\newcommand{\E}{\mathbb{E}}
\newcommand{\F}{\mathcal{F}}

\newcommand{\Y}{\mathcal{Y}}

\renewcommand{\L}{\mathcal{L}}

\renewcommand{\epsilon}{\varepsilon}
\renewcommand{\phi}{\varphi}
\newcommand{\eps}{\varepsilon}

    \def\independenT#1#2{\mathrel{\setbox0\hbox{$#1#2$}
    \copy0\kern-\wd0\mkern4mu\box0}}

\begin{document}

\title{Particle filtering in high-dimensional chaotic systems}

\author[N. Lingala]{Nishanth Lingala$^1$}
\address{$^1$Department of Aerospace Engineering\\
University of Illinois at Urbana-Champaign\\
306 Talbot Laboratory, MC-236\\
104 South Wright Street\\
Urbana, Illinois  61801, USA}
\email{lingala1@illinois.edu \\navam@illinois.edu \\hyeong2@illinois.edu}

\author[N.S. Namachchivaya]{N. Sri Namachchivaya$^1$}

\author[N. Perkowski]{Nicolas Perkowski$^2$}
\address{$^2$Institut f\"{u}r Mathematik\\
Humboldt-Universit\"{a}t zu Berlin\\
Rudower Chaussee 25\\
12489 Berlin\\
Germany}
\email{perkowsk@math.hu-berlin.de}

\author[H.C. Yeong]{Hoong C. Yeong$^1$}


\maketitle

\begin{abstract}
We present an efficient particle filtering algorithm for multiscale systems, that is adapted for simple atmospheric dynamics models which are inherently chaotic.
Particle filters represent the posterior conditional distribution of the state variables by a collection of particles, which evolves and adapts recursively as new information becomes available. The difference between the estimated state and the true state of the system constitutes the error in specifying or forecasting the state,
which is amplified in chaotic systems that have a number of positive Lyapunov exponents.
The purpose of the present paper is to show that the homogenization method developed in~\citet{INPY}, which is applicable to high dimensional multi-scale filtering problems,
along with important sampling and control methods can be used as a basic and flexible tool for the construction of  
the proposal density inherent in particle filtering. 
Finally, we apply the general homogenized particle filtering algorithm developed here to the 
Lorenz'96 (\citet{lorenz96}) atmospheric model that mimics mid-latitude atmospheric dynamics with microscopic convective processes.
\end{abstract}


\section{Introduction}

The main goal of filtering is to obtain, recursively in time, the best statistical estimate of a natural or physical system based on noisy partial observations of the same.
More precisely, filtering problems consist of an unobservable signal process and an observation process that is a function of the signal corrupted 
by noise. 
This is given by the conditional distribution of the signal given the observation process.
This paper deals with real time filtering of chaotic signals from atmospheric models involving many degrees of freedom.
Since small errors in our estimate of the current state of a chaotic system can grow to have a major impact on the subsequent forecast,
better estimate of the signal is needed to make accurate predictions of the future state. This is obviously a problem with significant importance 
in real time filtering and prediction of weather 
and climate which involves coupled atmosphere-ocean system as well as the spread of hazardous plumes or pollutants governed by extremely complex flows.
A complete study of these complex systems with practical impact, involves models of extremely unstable, chaotic dynamical systems with several million degrees of freedom.

Proper and accurate climate  models can only be obtained by combining the models with data. 
Lower dimensional climate models require first the identification of slowly evolving climate modes and fast evolving non-climate modes. 
Contrary to standard initial value problems, we do not have access to the initial state of these dynamical systems.
Instead, we have a known initial probability density function for the initial state and treat the model states as realizations of a random variable.
Lower dimensional climate models also contain noise terms that account for the interaction of the resolved climate modes with the neglected non-climate modes. 
Hence these problems require efficient new algorithms to estimate the 
present and future state of the climate models, based upon corrupted, distorted, 
and possibly partial observations of the climate modes and fast evolving non-climate modes. While perfect determination of 
the state is impossible under these noisy observations, it may still be 
desirable to obtain probabilistic estimates of the state conditioned on the information that is available. 

The sheer number of calculations required in directly solving large-scale random
dynamical systems becomes computationally overwhelming. Hence, we consider
the Lorenz'96 model of type II (\citet{lorenz96}) with two time-scale simplified ordinary 
differential equation describing advection, damping and forcing of some (slow) resolved atmospheric variables being 
coupled to some (fast) sub-scale variables as a nontrivial example of an atmospheric dynamics model. 
Though the model considered in this paper is a simple chaotic ``toy model" of the atmosphere, which is an excellent test-bed for schemes that will be developed in Section~\ref{S:HHPF}, 
it has a more realistic dimension than the multitude of ``Lorenz 1963"-type studies (i.e., 360 dimensions rather than 3 dimensions).  
Even though this simple model is still a long way from the Primitive Equations that we need to eventually study, it is hoped that the filtering methods presented 
here can be ultimately adapted for the realistic models that need to be computationally solved for weather and climate predictions.

In the past decade, vast amount of data has been produced via various
sources from network-connected sensor arrays to satellites, all with a wide range of scale separations.
To deal with such an onslaught of data, it is necessary to have a new
framework capable of harnessing and processing these data with multiscale models.
Hence, a data assimilation scheme that can handle chaotic systems that are sensitive to initial conditions~(characterized by positive
Lyapunov exponents) and noisy multiscale observations is needed. 
The theory of nonlinear filtering forms the framework in our study for
the assimilation of data into multiscale models.  

Particle filters represent the posterior conditional distribution of the state variables by a collection of particles, which evolves and adapts recursively as new information becomes available. Hence, particle filtering provides a recursive procedure for estimating an evolving state from a noisy observation process.
Despite the general applicability and rigorous convergence results, particle filters have not been extensively used in state estimation of high dimensional problems, for example, weather prediction, as pointed out in \citet{snyd08}. This is due to the fact that a large number of particles is required and the particle filters suffer from particle degeneracy (see, for example, \citet{daum03,snyd08}) in high dimensional systems. In this paper, we combine our study of stochastic dimensional 
reduction and nonlinear filtering to provide a rigorous framework for developing an algorithm for computing
lower dimensional particle filters which are specifically adapted to the complexities of the underlying multiscale signal.  
When the rates of change of different model variables differ by orders of magnitude,
efficient data assimilation can be accomplished by constructing a 
particle filter approach for nonlinear filtering equations for the 
coarse-grained signal. 
In moderate dimensional problems, particle filters are an attractive alternative to numerical 
approximation of the stochastic partial differential equations (SPDEs) by finite difference or finite element methods. 
Our approach described in this paper, not only reduces the computational burden for real time applications
but also helps solve the problem of particle degeneracy.
First objective is to predict the self-contained description of the coarse-grained dynamics without fully resolving
the dynamics described in fast scales. In these problems, extracting coarse-grained dynamics is at the heart a problem of
weak convergence of stochastic processes, or more precisely weak convergence of the laws of Markov processes. 

We begin in Section~\ref{S:Formulation} by presenting the general formulation of the multiscale nonlinear filtering problem. 
Here we introduce the homogenized equations that were derived in~\citet{INPY} for the reduced dimension unnormalized filter.
In Section~\ref{S:Model} we present the model that was originally suggested by Lorenz (Lorenz'96), which incorporated
a pattern with convective scales and introduced a crude model of $K$ slow variables plus $JK$ fast variables that varies with 
two distinct time scales.
This is an excellent test-bed for data assimilation schemes that are developed in this paper.
Section \ref{S:HHPF} outlines a sequential particle filtering algorithm (the Sequential Importance Sampling method) and a numerical algorithm for 
dimensional reduction (the Heterogeneous Multiscale Method of \citet{eijn03, ewei05}).  These two methods are combined to give an efficient 
algorithm for particle filtering in a multiscale  environment. 
Finally, we present in Section~\ref{S:Application}, the results from several data assimilation experiments on 
the Lorenz'96 model and discuss future research directions based on ``adaptive'' or ``targeted'' observations and sensor placement.

\section{Formulation of multiscale nonlinear filtering problems}
\label{S:Formulation}

The results presented here are set within the context of slow-fast dynamical systems, where the rates of change of different variables differ by orders of magnitude. The effects of the multiscale signal and observation processes via the study of lower dimensional Zakai equations in a canonical way was presented in~\citet{INPY11, INPY}, where the convergence of the optimal filter to the homogenized filter is shown using backward stochastic differential equations (BSDEs) and asymptotic techniques.
This paper provided rigorous mathematical results that support the numerical algorithms  based on the idea that stochastically averaged models provide qualitatively 
useful results which are potentially helpful in developing inexpensive lower-dimensional filtering. 
For the reduced nonlinear model an appropriate form of particle filter can be a viable and useful scheme.
Hence, we present the numerical solution of the lower dimensional  stochastic partial differential equation derived here, as it is applied to several higher dimensional multiscale applications.

We will present the main result of \citet{INPY} here. Let $(\Omega, \F, (\F_t), \Q)$ be a filtered probability space that supports a standard Brownian motion $(V, W, B)\in\R^{k\times l\times d}$. We consider the signal process that is the solution of the two time scale stochastic differential equations (SDEs) driven by $(W,V)$:
\begin{subequations}
\begin{align}\label{eq: diffusion-slow}
   dX^\epsilon_t & = b(X^\epsilon_t, Z^\epsilon_t) dt + \sigma(X^\epsilon_t, Z^\epsilon_t) dV_t, \quad X^\epsilon_0= \xi\in\R^m \\ \label{eq: diffusion-fast}
   dZ^\epsilon_t & = \frac{1}{\epsilon}f(X^\epsilon_t, Z^\epsilon_t)dt + \frac{1}{\sqrt{\epsilon}}g(X^\epsilon_t, Z^\epsilon_t)dW_t, \quad Z^\epsilon_0 = \eta\in\R^n.
\end{align}
\end{subequations}
Here, $\epsilon << 1$ is the timescale separation parameter, so $X^\epsilon$ is the slow component and $Z^\epsilon$ is the fast component. $W$, $V$ and $B$ are independent of each other as well as the random initial conditions $\xi$ and $\eta$. The functions $f$, $g$, $b$, $\sigma$ are assumed to be Borel-measurable. 

Associated to the signal is the $d$-dimensional observation $Y^\epsilon$ that is perturbed by the Brownian motion $B$, given by
\begin{align*}
   Y^\epsilon_t = \int_0^t h(X^\epsilon_s, Z^\epsilon_s) ds + B_t,
\end{align*}
where $h$ is a Borel-measurable function. Define the sigma algebra generated by the observation, $\Y^\epsilon_t = \sigma(Y^\epsilon_s: 0 \le s \le t)\vee \mathcal{N}$, where $\mathcal{N}$ are the $\Q$-negligible sets. 

The main objecive of filtering theory is to obtain the best estimate of the signal process based on information from the observation. The best estimate, called the optimal filter (the Zakai equation), is a conditional expectation that satisfies a recursive equation driven by the observation (\citet{Zakai1969}). For a formal definition, denote the optimal filter for the multiscale system by $\pi^\epsilon$, a finite measure on $\R^{m+n}$ . The goal is to calculate the (normalized) observation-dependent filter $\pi^\epsilon_t(\phi) \deff \E_\Q[\phi(X^\epsilon_t, Z^\epsilon_t) | \Y^\epsilon_t]$, where $\phi$ is a bounded measurable function on $\R^{m+n}$. 

In order to calculated the specified conditional expectation, it is easier to work on a new probability space $\P^\epsilon$, on which the observation $Y^\epsilon$ is a Brownian motion independent of the signal noise $(W,V)$. $\P^\epsilon$ is related to the original probability space $\Q$ by the Girsanov transform \citet{Bain2009}
\begin{align}\label{eq: Girsanov}
   D^\epsilon_t \deff \left.\frac{d\P^\epsilon}{d\Q}\right|_{\F_t} = \exp\left( - \int_0^t h(X^\epsilon_s, Z^\epsilon_s)^T dB_s - \frac{1}{2} \int_0^t |h(X^\epsilon_s, Z^\epsilon_s)|^2 ds\right),
\end{align}
which effectively removes the drift $h$ from the observation equation. Define an unnormalized filter as $\rho^\epsilon_t(\phi) \deff \E_{\P^\epsilon} [ \phi(X^\epsilon_t, Z^\epsilon_t)\left(D^\epsilon_t\right)^{-1}  |\Y^\epsilon_t ]$. The normalized and unnormalized filters are related using the Girsanov transform:
\begin{align*}
	\pi^\epsilon_t(\phi) = \E_\Q[\phi(X^\epsilon_t, Z^\epsilon_t) | \Y^\epsilon_t] 
	= \frac{\E_{\P^\epsilon}[\phi(X^\epsilon_t, Z^\epsilon_t) (D^\epsilon_t)^{-1} | \Y^\epsilon_t]}{\E_{\P^\epsilon}[(D^\epsilon_t)^{-1} | \Y^\epsilon_t]} = \frac{\rho^\epsilon_t(\phi)}{\rho^\epsilon(1)}.
\end{align*}

The unnormalized filter $\rho^\epsilon$ satisfies the Zakai equation (see, for example, \citet{Bain2009}):
\begin{align*}
   d\rho^\epsilon_t(\phi) = \rho^\epsilon_t(\L^\epsilon \phi) dt + \rho^\epsilon_t (h \phi) dY^\epsilon_t, \quad
   \rho^\epsilon_0(\phi) = \E_\Q[\phi(X^\epsilon_0, Z^\epsilon_0)].
\end{align*}
Here, $\L^\epsilon = \frac{1}{\epsilon}\L_F + \L_S$ is the differential operator associated to $(X^\epsilon, Z^\epsilon)$, with
\begin{align*}
   \L_F & = \sum_{i=1}^n f_i(x,z) \frac{\partial}{\partial z_i } + \frac{1}{2} \sum_{i,j=1}^n (g g^T)_{ij}(x,z) \frac{\partial^2}{\partial z_i  \partial z_j } \\
   \L_S & =  \sum_{i=1}^m b_i(x,z) \frac{\partial}{\partial x_i} +  \frac{1}{2} \sum_{i,j=1}^m (\sigma\sigma^T)_{ij}(x,z) \frac{\partial^2}{\partial x_i \partial x_j}
\end{align*}
where $\cdot^T$ denotes the transpose of a matrix or a vector.

Next, we briefly present a result from stochastic averaging theory for the multiscale setting of the problem considered. We assume that for every $x\in \R^m$, the solution $Z^x$ of \eqref{eq: diffusion-fast} with $X^\epsilon=x$ fixed is ergodic and converges rapidly to its unique stationary distribution $p_\infty(x:\cdot)$. In this case, it is well known that $X^\epsilon$ converges in distribution to a diffusion $X^0$ governed by an SDE
\begin{align*}
   dX^0_t = \bar{b}(X^0_t) dt + \bar{\sigma}(X^0_t) dV_t
\end{align*}
for appropriately averaged $\bar{b}$ and $\bar{\sigma}$. In other words, a stochastically averaged model provides a qualitatively useful approximation to the actual multiscale system. Hence, if we are only interested in estimating the {\it slow} process, or the ``coarse-grained'' dynamics, then we should make use of the homogenized diffusion $X^0$. Specifically, if we are only interested in the $x$-marginal, $\pi^{\epsilon,x}$, of the optimal $\pi^\epsilon$, then this $X^0$ can be used to construct an averaged, or homogenized, filter $\pi^0$ that can appropriately replace $\pi^{\epsilon, x}$. In high-dimensional problems, $\pi^0$ would be preferable to $\pi^{\epsilon,x}$ since calculation of $\pi^0$ using $X^0$ does not directly involve calculations of the fast process. 

By making use of $X^0$, we would like to find a homogenized (unnnormalized) filter $\rho^0$ that satisfies
\begin{align*}
   d\rho^0_t(\phi) = \rho^0_t(\bar{\L} \phi) dt + \rho^0_t(\bar{h}\phi) dY^\epsilon_t, \quad
   \rho^0_0(\phi) = \E_\Q[\phi(X^0_0)],
\end{align*}
such that for small $\epsilon$, the $x$-marginal of $\rho^\epsilon$, $\rho^{\epsilon, x}$, is close to $\rho^0$. We let the generator $\bar{\L}$ of $X^0$ be defined as
\begin{align*}
   \bar{\L} = \sum_{i=1}^m \bar{b}_i(x) \frac{\partial}{\partial x_i} + \frac{1}{2} \sum_{i,j=1}^m \bar{a}_{ij}(x,z) \frac{\partial^2}{\partial x_i \partial x_j}
\end{align*}
where $\bar{b}(x) = \int b(x,z) p_\infty(x,dz)$ and $\bar{a} = \int (\sigma\sigma^T)(x,z) p_\infty(x,dz)$. Also, define $\bar{h}(x) = \int h(x,z) p_\infty(x,dz)$. Note that the homogenized filter is still driven by the real observation $Y^\epsilon$ and not by a ``homogenized observation'', which is practical for implementation of the homogenized filter in applications since such avearaged observation is usually not available. However, even if such homogenized observation is available, using it would lead to loss of information for estimating the signal compared to using the actual observation.

Now define the measure-valued processes $\pi^0$ and $\pi^{\epsilon, x}$ in terms of $\rho^0$ and $\rho^{\epsilon, x}$ as $\pi^\epsilon$ was in terms of $\rho^\epsilon$:
\begin{align*}
  \pi^0_t (\phi) = \frac{\rho^0_t(\phi)}{\rho^0_t(1)} \qquad \text{and} \qquad \pi^{\epsilon, x}_t(\phi) = \frac{\rho^{\epsilon, x}_t(\phi)}{\rho^{\epsilon,x}_t(1)}.
\end{align*}
The main result of \citet{INPY} is that under appropriate assumptions on the coefficients of \eqref{eq: diffusion-slow}, \eqref{eq: diffusion-fast} and $h$, there exists a metric $d$ on the space of probability measures such that for every $T \ge 0$ there exists $C>0$ such that
\begin{align*}
	\E_\Q\left[ d(\pi^{\epsilon,x}_T, \pi^0_T)\right]\le \sqrt{\epsilon} C, ~\textrm{i.e.}~ \lim_{\epsilon\to 0} \E_\Q\left[ d(\pi^{\epsilon,x}_T, \pi^0_T)\right] = 0 ~\textrm{for any}~ T>0.
\end{align*}
In other words, the $x$-marginal of the optimal filter $\pi^\epsilon$ converges to the averaged filter $\pi^0$ in the space of probability measures as the separation parameter $\epsilon$ goes to $0$. Hence, the homogenized filter is an appropriate measure to use in place of the actual $\pi^{\epsilon,x}$ in estimating the ``coarse-grained'' dynamics $X^\epsilon$ in a setting with wide timescale separation. In terms of filtering applications, $\pi^0$ presents the advantage of not requiring exact knowledge of the fast dynamics for the purpose of estimating the ``coarse-grained'' dynamics. Only knowledge of the invariant measure of $Z^x$ is required, hence by applying appropriate multiscale averaging numerical schemes, computation and information storage for the fast dynamics can be reduced. 

The convergence of the normalized filters, $\pi^{\epsilon,x}\to\pi^0$, was shown by first obtaining the convergence of the unnormalized filters, $\rho^{\epsilon,x}$ to $\rho^0$. The dual representations of $\rho^{\epsilon,x}_T(\phi)$ and $\rho^{0}_T(\phi)$ were introduced as in \citet{Pardoux1979}:
\begin{align*}
   v^{\epsilon,T,\phi}_t(x,z) \deff \E_{\P^\epsilon_{t,x,z}}[\phi(X^\epsilon_T) \left(D^\epsilon_{t,T}\right)^{-1} | \Y^\epsilon_{t,T}], \quad \textrm{and} \quad v^{0,T,\phi}_t(x) = \E_{\P^0_{t,x}}[\phi(X^0_T) \left(D^0_{t,T}\right)^{-1} | \Y^\epsilon_{t,T}].
\end{align*}
$\P_{t,x,z}^\epsilon$ and $\P^0_{t,x}$ are the respective measures under which $(X^\epsilon,Z^\epsilon)$ and $X^0$ are governed by the same dynamics as under $\P^\epsilon$ and $P^0$, but $(X^\epsilon, Z^\epsilon)$ and $X^0$ stays in $(x,z)$ and $x$ until time $t$. $D^\epsilon_{t,T}$ and $D^0_{t,T}$ were defined as the Girsanov transform \eqref{eq: Girsanov}, but with moving limit of integration $t$,
\begin{align*}
   D^\epsilon_t \deff \left.\frac{d\P^\epsilon}{d\Q}\right|_{\F_t} = \exp\left( - \int_t^T h(X^\epsilon_s, Z^\epsilon_s)^T dB_s - \frac{1}{2} \int_t^T |h(X^\epsilon_s, Z^\epsilon_s)|^2 ds\right),
\end{align*}
\begin{align*}
   D^0_{t,T} = \exp\left( - \int_t^T \bar{h}(X^0_r)^T dY^\epsilon_r + \frac{1}{2} \int_t^T |\bar{h}(X^0_r)|^2dr \right),
\end{align*}
and $\Y^\epsilon_{t,T} = \sigma(Y^\epsilon_r - Y^\epsilon_t : t \le r \le T) \vee \mathcal{N}$ is the filtration generated by the observation over $[t,T]$, minus observation history from $0$ to $t$. The Markov property of $(X^\epsilon, Z^\epsilon, X^0)$ results gives the relations between the unnormalized filters and the respective duals: 
\begin{align}\label{eq: rho-v relations}
   \rho^{\epsilon,x}_T(\phi) = \int v^{\epsilon, T, \phi}_0 (x,z) \Q_{(X^\epsilon_0,Z^\epsilon_0)}(dx,dz) \quad \textrm{and} \quad \rho^0_T(\phi) = \int v^{0, T, \phi}_0 (x) \Q_{X^0_0}(dx).
\end{align}
Note that $\Q_{X^0_0} = \Q_{X^\epsilon_0}$, because the homogenized process has the same starting distribution as the un-homogenized one.

For fixed $T$ and $\phi \in C^2_b(\R^m, \R)$, we will write $v^\epsilon_t = v^{\epsilon, T, \phi}_t$ and $v_t^0 = v^{0, T, \phi}_t$. The dual process $v_t^\epsilon$ essentially represents the conditional expectation $\rho_T^{\epsilon,x}$ by an alternate conditional expectation that is run backwards in time from $T$ to $0$. For fixed starting point $(x,z)$ at $t=0$, this backward in time conditional expectation is constructed using processes $(X_T^\epsilon,Z_T^\epsilon)$ that started at $(x,z)$ and ran backwards to $t=0$. By integrating $v_0^\epsilon$ over $\Q_{(X_0^\epsilon,Z_0^\epsilon)}$, we are integrating over all possible starting points $(x,z)$, hence giving $\rho_T^\epsilon$. This interpretation is similar for $v^0$ of $\rho^0$. 

From \eqref{eq: rho-v relations}, we have
\begin{align}\label{eq: rhoepsilon minus rho0}
  \E[|\rho^{\epsilon,x}_T(\phi) - \rho^0_T(\phi)|^p] \le \int  \E[ | v^\epsilon_0(x,z) - v_0^0(x)|^p] \Q_{(X^\epsilon_0, Z^\epsilon_0)}(dx, dz).
\end{align}
So, if $| v^\epsilon_0(x,z) - v_0^0(x)|$ is small, then $|\rho^{\epsilon,x}_T(\phi) - \rho^0_T(\phi)|$ will also be small as long as $\Q_{(X^\epsilon_0, Z^\epsilon_0)}$ is well behaved and \eqref{eq: rhoepsilon minus rho0} will lead to the normalized filter convergence result. Therefore, the aim is to show that for nice test functions $\phi$, $\E[|v^\epsilon_0(x,z) - v^0_0(x)|^p]$ is small. 

The key point is that $v^\epsilon$ and $v^0$ solve backward SPDEs. We formally expand $v^\epsilon$ as
\begin{align*}
      v^\epsilon_t(x, z) = \underbrace{u^0_t(x,z)}_{v^0(t,x)} + \underbrace{\epsilon u^1_{t/\epsilon} \left( x, z \right)}_{\psi(t,x,z)} + \underbrace{\epsilon^2 u^2_{t/\epsilon}(x, z)}_{R(t,x,z)}.
\end{align*}
Then, $v^0$, $\psi$ and $R$ satisfy linear partial differential equations with appropriate terminal conditions. By existence and uniqueness of the solutions to these \emph{linear} equations, we can apply superposition to obtain that indeed
\begin{align*}
   v^\epsilon_t(x,z) = v^0_t(x) + \psi_t(x,z) + R_t(x,z),
\end{align*}
where $\psi$ and $R$ are the corrector and remainder terms, respectively. Based on the expansion, the problem of showing $L^p$-convergence of $v^\epsilon$ to $v^0$ reduces to showing $L^p$-convergence of $\psi$ and $R$ to $0$. 

Details of the proof are provided in \citet{INPY}. The outline of the method of proof is as follows: The backwards SPDEs 
were converted to their respective proabilistic representations, which are backward doubly stochastic differential equations (BDSDEs). The diffusion operators were replaced by the associated diffusions and explicit estimates for the finite dimensional BDSDEs in terms of the transition density function of the fast process were able to be obtained. ~\citet{Pardoux2003} proved very precise estimates for this transition function, and were used to obtain the desired bounds on $\psi$  and $R$.

\section{The Lorenz'96 System}
\label{S:Model}

The systematic strategy for the identification of slowly evolving climate modes and fast evolving non-climate modes requires a lengthy analysis.
In this paper, we consider a simple atmospheric model, which nonetheless exhibits many of the difficulties arising in realistic models, to gain insight into predictability and data assimilation.
The Lorenz'96 model was originally introduced (in \citet{lorenz96}) to mimic multiscale mid-latitude weather, considering an unspecified scalar meteorological quantity at $K$ equidistant grid points along a latitude circle:
\begin{align}
	\label{eq: lorenz-slow}
	&\dot{X}_t^k = -X_t^{k-1}(X_t^{k-2}-X_t^{k+1}) - X_t^k + F_x +\frac{h_x}{J}\sum_{j=1}^{J}Z_t^{k,j}, \quad k = 1,\ldots,K, \\
	\label{eq: lorenz-fast}
	&\dot{Z}_t^{k,j} = \frac{1}{\epsilon}\left\{-Z_t^{k,j+1}(Z_t^{k,j+2}-Z_t^{k,j-1}) - Z_t^{k,j} + h_zX_t^k\right\}, \quad j = 1,\ldots,J. 
\end{align}
Equation \eqref{eq: lorenz-slow} describes the dynamics of some atmospheric quantity $X$, and $X_t^k$ can represent the value of this variable at time $t$, in the $k^{\rm th}$ sector defined over a latitude circle in the mid-latitude region ({\it Note:} We use superscripts $k$ and $j$ to conform with the typical spatial indexing notation used for the Lorenz '96 model. In sections that follow, subscripts $k$ and $j$ will be used as discrete time indices, not to be confused with the spatial indices of the Lorenz model). 
The latitude circle is divided into $K$ sectors (with values of $K$ ranging from $K=4$ to $K=36$). Each $X_t^k$ is coupled to its neighbors $X_t^{k+1}$, $X_t^{k-1}$, and $X_t^{k-2}$ by \eqref{eq: lorenz-slow}. \eqref{eq: lorenz-slow} applies for all values of $k$ by letting $X_t^{k+K} = X_t^{k-K} = X_t^{k}$, so, for example for $k=1$, $X_t^{k+1} = X_t^{2}$, $X_t^{k-1} = X_t^{36}$, and $X_t^{k-2} = X_t^{35}$. 

The system was extended to study the influence of multiple spatio-temporal scales on the predictability of atmospheric flow by the dividing each segment $k$ into $J$ subsectors~($J=4$ to $J=10$), 
and introducing a fast variable, $Z_t^{k,j}$ given by~\eqref{eq: lorenz-fast}, associated with each subsector. 
Thus, each $X_t^{k}$ represents a slowly-varying, large amplitude atmospheric quantity, with $J$ fast-varying, 
low amplitude, similarly coupled quantities, $Z_t^{k,j}$, associated with it.
In the context of climate modeling, the slow component is also known as the resolved climate modes while the rapidly-varying component is known as the unresolved non-climate modes.
The coupling terms between neighbors model advection between sectors and subsectors, while the coupling between 
each sector and its subsectors model damping within the model. The model is subjected to linear external forcing, $F_x$, on the slow timescale. 

The dynamics of the unresolved modes can be modified to include nonlinear self-interaction effects by adding forcing in the form of stochastic 
terms (see, for example, \citet{Majda2001, Majda2003}). The use of stochastic terms to represent nonlinear self-interaction effects at short timescales in the 
unresolved modes is appropriate if we are only interested in the coarse-grained dynamics occuring in the long timescale. This is called stochastic 
consistency in \citet{Majda2003}. Considering \eqref{eq: lorenz-fast}, where only quadratic nonlinearity is present, the motivation behind adding 
stochastic forcing is thus to model higher order self-interaction effects. This can be done using a mean-zero Ornstein-Uhlenback process, 
specifically a 
process with amplitude $\frac{1}{\sqrt{\epsilon}}$, that is,
\begin{align}
	\label{eq: lorenz-fast-noise}
	&\dot{Z}_t^{k,j} = \frac{1}{\epsilon}\left\{-Z_t^{k,j+1}(Z_t^{k,j+2}-Z_t^{k,j-1}) - Z_t^{k,j} + h_zX_t^k\right\} + \frac{1}{\sqrt{\epsilon}} \zeta(t).
\end{align}
This is done more explicitly in a general context in Section \ref{S:HHPF}.  
The two-scale Lorenz'96 model was also used extensively by \citet{Wilks2005} to study stochastic parametrization and
by \citet{Lorenz1998} for analyzing targeted observations, and by~\citet{Herrera11} and
several others for analyzing the influence of large-scale spatial patterns on the growth of small perturbations.

\section{Homogenized Hybrid Particle Filter (HHPF)}
\label{S:HHPF}
Numerical simulation of multiscale dynamical systems is quite problematic because of the wide separation in the time-scales involved.  
Based on the results presented in Section \ref{S:Formulation}, we combine our study of stochastic dimensional reduction and nonlinear filtering to provide a rigorous framework for developing an algorithm for lower dimensional particle filters specifically adapted to the complexities of underlying multiscale signals.  
The proposed particle filter algorithm is adapted to a system with time-scales separation by incorporating a homogenization scheme in the nonlinear filter, in conjunction with the stochastic dimensional reduction results of Section \ref{S:Formulation}.

In order to apply stochastic dimensional reduction in the problem presented in Section \ref{S:Model}, we introduce a stochastic forcing term in \eqref{eq: lorenz-fast} to represent external forcing effects on the system due to unresolved dynamics.   
To illustrate the homogenization scheme, we consider a general system of stochastic differential equations (SDEs) that corresponds with the problem of Section \ref{S:Model}, with stochastic forcing in the fast component:
\begin{align}\label{eq: multiscale-model-HMM}
  \dot X^\eps_t &= b(X^\eps_t,Z^\eps_t), \quad X^\eps_0 \sim \mathcal{N}(\mu^x_0, \sigma^x_0) \\ \nonumber
  \dot Z^\eps_t &= \eps^{-1}f(X^\eps_t,Z^\eps_t) 
  +\eps^{-1/2} g(X^\eps_t,Z^\eps_t)\dot W_t, \quad Z^\eps_0 \sim \mathcal{N}(\mu^z_0, \sigma^z_0). 
\end{align}

In the following two subsections we explain the Heterogeneous Multiscale Method and Sequential Importance Sampling. These techniques are then combined in the Homogenized Hybrid Particle Filter.

\subsection{Multiscale numerical integration}
\label{s:HMM}
For numerical simulations of \eqref{eq: multiscale-model-HMM}, the timestep $\delta t$
required for the forward integration of the fast process $Z^\eps$ needs to be
smaller than $\eps$ for stability of the integration scheme. However, with such timestep, significant changes in the slow variable can only be seen on the time-scale 
of $\mathcal{O}(1)$, i.e., much of the computational resources is wasted in solving for an almost stationary evolution of the slow process $X^\eps$. 

Based on the results presented in Section~\ref{S:Formulation}, we can solve this problem by adopting a reduced system through stochastic averaging, that is,
\begin{equation}\label{eq: slow-averaged-ch4}
  \dot{\bar{X}}_t = \bar b(\bar X_t),
\end{equation} 
where
\begin{equation}\label{eq: drift-averaged-ch4}
  \bar b(x) = \int b(x,z) \mu_x(dz).
\end{equation} 
By adopting \eqref{eq: slow-averaged-ch4} with \eqref{eq: drift-averaged-ch4}, we are under the assumption that \eqref{eq: multiscale-model-HMM} is exponentially mixing, and the Doeblin condition is satisfied. Stated informally:
With the slow process fixed at $X_t^\eps = x$, the corresponding fast process $Z^{x}$ has a unique invariant measure $\mu_x(dz)$, 
i.e. the transition probability measure $P^{x}(z,t)$ of $Z_t^{x}$ converges exponentially to $\mu_x(dz)$ in the weak sense as $t\to \infty$, locally uniformly in $x$ and $z$. This implies that for any test function $\varphi \in C_b(\R^m)$ 
\begin{align*}
\E_z\left[ \varphi (Z_t^{x}) \right] \to \int \varphi (z)\mu_x(dz) \quad \rm{as} \quad t\to \infty,
\end{align*}
uniformly in $x$ and $z$ in any compact set.

However, the evaluation of the high dimensional integration in~\eqref{eq: drift-averaged-ch4} is nontrivial, and usually it is impossible to obtain an invariant distribution of the fast variable analytically. To determine the invariant distribution numerically, we adopt the Heterogeneous Multiscale Method (HMM) introduced in \citet{eijn03}. The HMM is a method of determining the effective dynamics \eqref{eq: slow-averaged-ch4} through numerical approximation of the invariant distribution of the fast process. It is based on the observation that the fast variable $Z^\eps$ reaches its invariant distribution (equilibrium) on a time-scale much smaller than the time-scale needed to evolve the slow variable $X^\eps$ (the Doeblin condition). 
This implies that  we could use a much larger timestep value $\Delta t$ for the evolution of the slow process while keeping the timestep for the fast process small for stability.  

Following the procedure presented in \citet{eijn03}, we describe the HMM. The evolution of the averaged equation \eqref{eq: slow-averaged-ch4} is
approximated numerically using a forward integration scheme. For simplicity, we will use the Euler and Euler-Maruyama schemes in the following description, although higher order schemes are also applicable (see, for example, \citet{kloe92}). Consider the interval $[0,T]$ discretized into timesteps of size $\Delta t = \left\lfloor \frac{T}{N} \right\rfloor$. Let $t_k \deff k\Delta t$ and write $X_{t_k}$ as $X_k$.
\begin{equation}\label{eq: macro-solver}
  \bar{X}_{k+1} = \bar{X}_k + \tilde b(\bar{X}_k) \Delta t
\end{equation}  
is the \textit{macro-solver} with $\Delta t$ being the \textit{macro-timestep} ({\it Note:} As mentioned in Section \ref{S:Model}, subcript $k$ here indicates time index, different from the spatial index superscript $k$ in the Lorenz '96 model).
The value of $\tilde b$, which is an approximation of the averaged coefficient $\bar b$ in \eqref{eq: drift-averaged-ch4}, can be calculated as 
\begin{equation}\label{eq: MC-averaged}
  \tilde b(\bar{X}_k) = 
  \frac{1}{MN_m} \sum_{r=1}^M \sum_{j = n_T}^{n_T+N_m} b(\bar{X}_k,Z^{\epsilon,r}_{k,j})
\end{equation} 
where $M$ is the number of replicas of the fast process $Z$  for spatial averaging, $N_m$ is the number of \textit{micro-timesteps} $\delta t$ for time averaging, and $n_T$ is the number of micro-timesteps skipped to eliminate transient effects. The evolution of $Z^{\epsilon,r}_{k,j}$ is
governed by the following \textit{micro-solver}
\begin{equation}\label{eq: micro-solver}
  Z^{\epsilon,r}_{k,j+1} = Z^{\epsilon,r}_{k,j} + \frac{1}{\epsilon} f(\bar{X}_k,Z^{\epsilon,r}_{k,j}) \delta t + \frac{1}{\sqrt{\epsilon}} g(\bar{X}_k,Z^{\epsilon,r}_{k,j}) \delta W.
\end{equation} 
Note that a combination of spatial and temporal averaging is used in \eqref{eq: MC-averaged} but it is shown in \citet{eijn03} that the combinations of $M$, $N_m$, and $\delta t$ can be chosen based on error analysis such that no spatial averaging or no spatial and temporal averaging is required. For more detailed explanation and error analysis, refer to the references \citet{eijn03}, \citet{fatkullin04}.


\subsection{Homogenized Hybrid Particle Filter (HHPF)}
\label{s:HHPF}
\noindent 
The algorithm for the continuous-time HHPF is presented in \citet{Park2010}.
As in standard particle filtering methods, the continuous-time equations can be discretized and the filtering can be done on the resulting discrete-time models. 
Here, we present the discrete-time version of the HHPF using the sequential importance sampling (SIS) algorithm, which is also commonly known as bootstrap filtering (see, for example, \citet{arulam02}, \citet{gord93}). 

We will first provide a brief overview of the idea behind importance sampling and then illustrate how it is applied sequentially in particle filters. Then we present how the HHPF uses the SIS algorithm.

\subsubsection{Importance Sampling}\label{sbsbsec:ImpSamp}
In particle filtering, there is always the need to represent some distribution using a collection of particles.  When it is difficult to sample from a given distribution, the idea is to sample from another distribution that is more tractable to sample from, and properly use that sample to represent the distribution of interest.

Importance sampling is a technique for approximating integrals with respect to one probability distribution using a collection of samples from another. Let $p$ be the target distribution of interest over space $\mathbb{X}$ and $q \gg p$ ($q$ is absolutely continuous with respect to $p$) be the distribution from which sampling is done ($q$ is also called the proposal distribution). Denote by $\E_p [.]$ and $\E_q [.]$ the expectation with respect to the distributions $p$ and $q$, respectively. For any integrable function $ \varphi :\mathbb{X}\to\R$, we have
\begin{align}\label{eq: importance-sampling}
	\E_p\left[ \varphi(X) \right] & = \int_\mathbb{X} \varphi(x)p(dx) = \int_\mathbb{X} \varphi(x)\frac{dp}{dq}(x) q(dx) \\ \nonumber 
	& = \int_\mathbb{X} \varphi(x)w(x)q(dx) = \E_q\left[ w(X)\varphi(X) \right],
\end{align}
where $w\deff \frac{dp}{dq}$.

A collection $\{x^i\}_{i=1}^{N_s}$ of $N_s$ particles can be sampled from $q$ and the particles can be weighted according to $w^i\propto \frac{dp}{dq}(x^i)$ to represent the target distribution $p$ i.e.
$$p(x)\approx \sum_{i=1}^{N_s}{w}^i\delta(x-x^i).$$
The weights ${w}^i$ are normalized such that $\sum {w}^i=1$.
 
The strong law of large numbers can be employed to verify that the empirical average of $\varphi$ with respect to the weighted sample from $q$ converges as $N_s\to \infty$, with probability 1, to the expected value of $\varphi$ under the target distribution $p$, i.e. 
$$\int \left[\sum_{i=1}^{N_s}{w}^i\delta(x-x^i)\right]\varphi(x)dx=\sum_{i=1}^{N_s}{w}^i\varphi(x^i)\to \E_p\left[ \varphi(X) \right].$$

\subsubsection{Sequential Importance Sampling (SIS)} \label{ss:SIS}
The SIS algorithm is a technique of using Monte-Carlo simulations for Bayesian filtering through the incorporation of importance sampling. 
Consider a discrete-time signal $X_{t_k}$ 
and a discrete-time observation $Y_{t_k}$ and let $p\left(\left. x_{0:k}\right| y_{0:k}\right)$ be the density of the target posterior distribution at timestep $t_k$. The SIS algorithm approximates the target distribution using appropriately weighted samples from a proposal density $q\left( \left. x_{0:k} \right| y_{0:k}\right)$.

Suppose we represent $p\left(\left. x_{0:k}\right| y_{0:k}\right)$ using a collection $\{x^i_{0:k}\}_{i=1}^{N_s}$ of $N_s$ particles sampled according to $q\left( \left. x_{0:k} \right| y_{0:k}\right)$ as 
$$p\left( \left. x_{0:k} \right| y_{0:k}\right) \approx \sum_{i=1}^{N_s} w^i_k \delta (x - x^i_{0:k}),$$
where $w^i_k \propto \frac{p\left( \left. x^i_{0:k} \right| y_{0:k}\right)}{q\left( \left. x^i_{0:k} \right| y_{0:k}\right)}$, with $\sum_i w^i_k=1$. 
We choose to arrive at proposal densities sequentially
\begin{align*}
q\left( \left. x_{0:k} \right| y_{0:k}\right) & = q\left( \left. x_k \right| x_{k-1},y_{k}\right) q\left( \left. x_{0:k-1} \right| y_{0:k-1}\right).
\end{align*}
Making use of the identity (see, for example, \citet{arulam02}, equation (45)) 
\begin{align*}
	p\left( \left. x_{0:k} \right| y_{0:k}\right) & = \frac{ p\left( \left. y_k \right| x_k\right) p\left( \left. x_k \right| x_{k-1}\right) p\left( \left. x_{0:k-1} \right| y_{0:k-1}\right) }{ p\left( \left. y_k \right| y_{0:k-1}\right) },
\end{align*}
we write (after removing $p\left( \left. y_k \right| y_{0:k-1}\right)$ because it is same for all the particles)
\begin{align}\label{eq: sequential-weights}
w^i_k & \propto \frac{p\left( \left. y_k \right| x_k^i\right) p\left( \left. x^i_k \right| x^i_{k-1}\right) }{q\left( \left. x^i_k \right| x^i_{k-1},y_{k} \right) } \frac{p\left( \left. x^i_{0:k-1} \right| y_{0:k-1}\right)}{q\left( \left. x^i_{0:k-1} \right| y_{0:k-1}\right)} \\ \nonumber
& \propto \frac{p\left( \left. y_k \right| x_k^i\right) p\left( \left. x^i_k \right| x^i_{k-1}\right) }{q\left( \left. x^i_k \right| x^i_{k-1},y_{k} \right) } w^i_{k-1}.
\end{align}
Hence we have a sequential formula for computing the unnormalized particle weights. Because our choice of $q$ is done sequentially, the proposal density depends on the location of the particle only at the previous time step. Only $x^i_{k-1}$ need be stored and the rest of the path $x^i_{0:k}$ can be discarded. We then have
$$p\left( \left. x_k \right| y_{0:k}\right) \approx \sum_{i=1}^{N_s} {w}^i_k \delta (x - x^i_{k})$$
with weights updated according to \eqref{eq: sequential-weights} and normalized by $\sum_i w^i_k=1$.

So, the SIS algorithm is this: 
At step $k-1$ the location of particles $\{x^i_{k-1}\}_{i=1}^{N_s}$ is known. New observation $y_k$ is recorded. Choose a form for the proposal $q(\cdot | x^i_{k-1},y_k)$. Sample a particle according to this proposal density and get the new location $x^i_k$. Evaluate the number $q(x^i_k | x^i_{k-1},y_k)$, and then evaluate the quantities in the numerator of equation \eqref{eq: sequential-weights}, using the sensor dynamics and the signal dynamics. Then update the weights according to equation \eqref{eq: sequential-weights}.

In filtering with a fixed number of particles, it is crucial to keep the variance of the weights to minimum: if a collection of particles is such that the weight is concentrated in a small number of particles, this collection represents the distribution poorly. One can better it by selecting a collection which has many particles near the region of high concentration and the particles sharing nearly equal weights. As can be seen from the weight update equation \eqref{eq: sequential-weights}, the particle weights depend crucially on the choice of the proposal distribution $q$. So, $q$ should be choosen with an aim of minimizing the variance of particle weights. 

It can be shown \citet{arulam02} that the proposal density which keeps the variance of the weights to a minimum is
\begin{align}\label{IS:whatisqopt}
q^{\mathrm{opt}} (x_{k}| x_{k-1},y_{k}) = \frac{p\left( \left. y_k \right| x_k\right) p\left( \left. x_k \right| x_{k-1}\right)}{\int p\left( \left. y_k \right| x_k\right) p\left( \left. x_k \right| x_{k-1}\right) dx_k}.
\end{align}
Employing this in weight update equation \ref{eq: sequential-weights} we have that, if we choose the optimal proposal density,
\begin{align}\label{IS:wtaccqopt}
w^i_k & \propto w^i_{k-1}\int p\left( \left. y_k \right| x_k\right) p\left( \left. x_k \right| x_{k-1}^i\right) dx_k.
\end{align}
In general case, it is difficult to sample from $q^{\mathrm{opt}}$. However, it is easy when both the likelihood $p\left( \left. y_k \right| x_k\right)$ and the conditional prior $p\left( \left. x_k \right| x_{k-1}\right)$ are Gaussians, cf. \citet{arulam02}.

Consider the system dynamics and observation equation:
\begin{align}\label{IS:example}
  X_{k+1}&=F(X_k) + \sigma_{X} W_{k+1} \\
 Y_{k+1}&=HX_{k+1}+\sigma_{Y}V_{k+1}
 \label{IS:example1}
\end{align}
where $W_k$ and $V_k$ are independent centered Gaussian increments with variance one and $Q \deff \sigma_X\sigma_X^T$, $R\deff \sigma_Y\sigma_Y^T$ are strictly positive definite. Since this subsection on SIS and the next one on optimal control discuss a general method, we have used $F$ to denote the vector field instead of $b$ and $f$ as in the previous sections. We then have that the likelihood $p(y_k|x_k)=\mathcal{N}(Hx_k,R)$ and the conditional prior $p\left( \left. x_k \right| x_{k-1}\right)=\mathcal{N}(F(X_{k-1}),Q)$. Using \eqref{IS:whatisqopt}, we have  
\begin{align}
\label{SIS:example_evolve_particles}
q^{\mathrm{opt}}(x_{k}| x_{k-1},y_{k})&=\mathcal{N}(F(x_{k-1})+\alpha(x_{k-1},y_k),\hat{Q}), \nonumber \\
\hat{Q}&=\left(Q^{-1}+H^TR^{-1}H\right)^{-1},  \nonumber \\
\alpha(x_{k-1},y_k)&=\hat{Q}H^TR^{-1}(y_k-HF(x_{k-1})).
\end{align}
This $q^{\mathrm{opt}}$ is a Gaussian. Once we have particle locations $\{x^i_{k-1}\}_{i=1}^{N_s}$ representing the posterior at time $k-1$, and the observation $y_k$ is recorded, we can evaluate $\alpha(x_{k-1},y_k)$. We can then sample a particle $x^i_{k}$ from the above Gaussian $\mathcal{N}(F(x_{k-1}^i)+\alpha(x_{k-1}^i,y_k),\hat{Q})$ and arrive at a collection of particles $\{x^i_{k}\}$. Alternatively, the particles can be evolved according to 
\begin{align}\label{IS:example_evolve_particles}
X_{k+1}&=F(X_k) + \alpha(X_k,y_{k+1}) + \hat{\sigma}_{X} W_{k+1}
\end{align}
where $\hat{\sigma}_{X}$ is such that $\hat{\sigma}_{X}\hat{\sigma}_{X}^T = \hat{Q}$. Then $X_{k}^i$ behaves like a particle sampled from $q^{\mathrm{opt}} (\cdot| x_{k-1}^i,y_k).$ The weights are updated according to \eqref{IS:wtaccqopt}: we will have
\begin{align}
w^i_k\,\,&\propto\,\, w^i_{k-1}\exp\left\{-\frac12 (y_k-HF(x_{k-1}^i))^T\hat{R}^{-1}(y_k-HF(x_{k-1}^i))\right\}, \nonumber \\
\hat{R}^{-1}&=R^{-1}\left(\mathbf{1}-H\hat{Q}H^TR^{-1}\right).
\end{align}

\subsubsection{Stochastic optimal control approach}
\label{ss:SOC}

We consider the same discrete-time nonlinear system with linear observation \eqref{IS:example}, \eqref{IS:example1}. 
The particle method presented in this section consists of control terms in the ``prognostic" equations, that nudge the particles 
toward the observations. We nudge the particles by applying a control $u_k(y,x)$ according to
\begin{align}\label{IS:example_nudge}
  X_{k+1}&=F(X_k) + u_k(y_{k+1},X_k)+ \sigma_{X} W_{k+1}.
\end{align}
The technique for determining the nudging term $u_k(y_{k+1},x)$ is by method of stochastic optimal control
by minimizing a quadratic cost:
\begin{align}\label{eq: cost}
  J := \E_{k,x}\frac12\left[  u_k^T(y,x)Q^{-1} u_k(y,x) + (y - h(X^{(k,x)}_{k+1}))^T R^{-1} (y-h(X^{(k,x)}_{k+1}))\right], 
\end{align}
where $Q$, $R$ are the signal and observation noise covariance matrices and by $X^{(k,x)}$ we mean that the process $X$ started at time $k$ at the value $x$. Again, we assume the linear sensor function with observation available at every timestep, given by \eqref{IS:example1}. 
The purpose of the present section is to show that control methods can be used as a basic and flexible tool for the construction of  
the proposal density inherent in particle filtering. 
In this framework, $u$ can be interpreted as the optimal control for minimizing the cost $J$. The first term in~\eqref{eq: cost} represents the control energy and if we allow $u$ to become too big, then heuristically all the particles will coincide with the observation. Then the particles will be a sample from a Dirac distribution, whereas the conditional distribution that we try to simulate is absolutely continuous. The second term in~\eqref{eq: cost} represents the distance between $HX_{k+1}$ and the observation that we want minimized.
Covariance matrices $Q$ and $R$ in the quadratic terms indicate that dimensions of the signal and observation that have larger noise variance are penalized less by the control. 
This means that in directions where noise amplitude is large, we allow for more correction by taking $Q^{-1}$, which puts less penalty on the size of the control in the dimensions with large noise amplitude.
Similarly, the terminal cost given by the second term in~\eqref{eq: cost} incurs a penalty for being far away from the actual signal based on observation, but in directions where the quality of the observation is not very good, we allow our particle to be further away from the observation, hence $R^{-1}$.

Define the value function
\begin{align*}
	V^{opt}(k,x) := \inf_{u_k(y,x)} \E_{k,x}\frac12\left[  u_k^T(y,x)Q^{-1} u_k(y,x) + (y - h(X^{(k,x)}_{k+1}))^T R^{-1} (y-h(X^{(k,x)}_{k+1}))\right].
\end{align*} 
Using \eqref{IS:example} and \eqref{IS:example1}, and substituting for $X^{(k,x)}_{k+1}$ in the second expression in the value function, we have
\begin{align*}
	& \left(y - HX^{(k,x)}_{k+1}\right)^TR^{-1}\left(y - HX^{(k,x)}_{k+1}\right) \\
	&  = \left(y - H\tilde{F}(x,u_k)\right)^TR^{-1}\left(y - H\tilde{F}(x,u_k)\right) - 2\left(y - H\tilde{F}(x,u_k)\right)^TR^{-1}\left(H{\sigma}_XW_{k+1}\right) \\
	& \quad + \left(H{\sigma}_XW_{k+1}\right)^TR^{-1}\left(H{\sigma}_XW_{k+1}\right),
\end{align*}
where $\tilde{F}(x,u_k) = F(x) + u_k(y,x)$. Because $u_k$ depends only on $y$ and $x$, both of which are given, we have $\E_{k,x}\left[u_kQ^{-1}u_k\right]=u_kQ^{-1}u_k$, 
and similarly 
\begin{align*}
	\E_{k,x}\left( \left(y - H\tilde{F}(x,u_k)\right)^TR^{-1}\left(y - H\tilde{F}(x,u_k)\right) \right) = \left(y - H\tilde{F}(x,u_k)\right)^TR^{-1}\left(y - H\tilde{F}(x,u_k)\right).   
\end{align*}
$W_{k+1}$ is a standard Gaussian random variable independent of $X_k=x$, so
\begin{align*}
	\E_{k,x}\left[ \left(y - H\tilde{F}(x,u_k)\right)^TR^{-1}\left(H{\sigma}_XW_{k+1}\right) \right] & = \left(y - H\tilde{F}(x,u_k)\right)^TR^{-1}H{\sigma}_X \E\left[ W_{k+1}\right]=0, 
\end{align*}
\begin{align*}
	\E_{k,x}\left[ \left(H{\sigma}_XW_{k+1}\right)^TR^{-1}\left(H{\sigma}_XW_{k+1}\right)\right] & =  \operatorname{tr}\left(\left(H{\sigma}_X\right)^TR^{-1}\left(H{\sigma}_X\right)\right)
\end{align*}
Therefore, 
\begin{align*}
	V(k,x) = \frac{1}{2}\inf_{u_k}\left\{ u_k^TQ^{-1}u_k + \left(y - H\tilde{F}(x,u_k)\right)^TR^{-1}\left(y - H\tilde{F}(x,u_k)\right) + \operatorname{tr}\left(\left(H{\sigma}_X\right)^TR^{-1}\left(H{\sigma}_X\right)\right) \right\}
\end{align*}
and 
\begin{align*}
	\frac{\partial}{\partial u_k}V(k,x) & = {Q}^{-1}u_k + H^TR^{-1}H(F(x)+u_k) - H^TR^{-1}y \\
	& = ({Q}^{-1} + H^TR^{-1}H)u_k - H^TR^{-1}\left(y-HF(x)\right),
\end{align*}
hence the optimal control is 
\[ u_k^{opt} = ({Q}^{-1} + H^TR^{-1}H)^{-1}H^TR^{-1}(y-HF(x)), \]
which is similar to the solution from the previous approach, with the difference being the noise term, $\hat{\sigma}_X$ in \eqref{IS:example_evolve_particles} and ${\sigma}_X$ in \eqref{IS:example_nudge}. 

This section provided the results related to the control design of the particles~(prior to updating the weights) that is needed to nudge the particle solutions 
toward the observations. This procedure consists of adding, forcing terms to the ``prognostic" equations  for the construction of the proposal density inherent in particle filtering. However, it is possible that the nudging terms may  become too large and destroy the balance between the effects of the noise and the control terms. The stochastic optimal control approach presented in this section is similar to the derivation of the 4D-VAR method that is used in geophysical data assimilation (see, for example, \citet{Kalnay2003}). The 4D-VAR method considers the problem of determining the best initial condition at time $t_0$ for the forward integration of the model PDEs based on discrete observations collected, up to a finite time $t_K$, in the future of $t_0$. In the 4D-VAR method, the cost function to be minimized with respect to the initial condition $x(t_0)$ is 
\begin{align*}
	J(x(t_0)) &=\frac12[x(t_0)-x^b(t_0)]^TB_0^{-1}[x(t_0)-x^b(t_0)] \\
	&\quad + \frac12[H(x(t_K))-y(t_K)]^TR^{-1}[H(x(t_K))-y(t_K)],
\end{align*}
where $x^b(t_0)$ was predicted using the model equations from time before $t_0$ and $x(t_K)$ is obtained by integration of the model PDEs using $x(t_0)$ as initial condition. From this point of view, the stochastic optimal control approach presented here can be viewed as determining the optimal initial condition at every discrete time $t_k$ using the next available observation at $t_{k+1}$. The optimal control $u_k^{opt}$ is the correction made to the state $x_k$ predicted from $t_{k-1}$.  

The SIS presented in Section~\ref{ss:SIS} allows the modification of the drift terms as well as the stochastic coefficients in the  ``prognostic" equations and SIS is an easy approach to implement numerically, therefore we will use SIS in conjunction with HHPF throughout this paper.
However, when dealing with sparse data, a proposal density based on stochastic control theory presented in this section is essential,  as can be shown in~\citet{LNPY}.

\subsubsection{HHPF}
\label{s:HHPF-alg}
In the following we describe how the HHPF uses the SIS algorithm. For now, we would choose the proposal density $q\left( \left. \cdot \right| x^i_{k-1},y_{k} \right)=p(\cdot|x^i_{k-1})$ i.e. the conditional prior which can be obtained from the signal dynamics. A particle can be sampled from this $q$ by propagating the location of the particle at $k-1$ using the signal dynamics.  We note that by choosing such a $q$ we are being blind to the observation $y_k$. We address a better choice in Section \ref{s:ISuseqopt}.

The HHPF is developed based on the results presented in Section \ref{S:Formulation}. It incorporates the HMM described in Section \ref{s:HMM} in the particle evolution step of a particle filtering algorithm. Although the problem presented in Section \ref{S:Model} has continuous-time signal, the HHPF is applied on the associated discretized model.
 
Consider the time interval $[0,T]$, discretized into macro-timesteps of size $\Delta t = \left\lfloor \frac{T}{N} \right\rfloor$ and denote the micro-timestep by $\delta t << \Delta t$, where $\delta t$ is chosen small enough compared to $\epsilon$ for numerical stability. 
In the context of the HMM, the intervals $[k \Delta t, (k+1) \Delta t]$, $k = 1,\ldots,N-1$, are the macro-timestep intervals over which evolution of the slow process occurs. The fast process is evolved over micro-timestep intervals $[j \delta t, (j+1) \delta t]$, $j=1,\ldots,n_T + N_m$ within each macro-timestep interval.
The discretized version of the equation \eqref{eq: slow-averaged-ch4} governing the coarse-grained dynamics $\bar X_t$, is used for propagation of particles $\left\{ \bar x^i_k \right\}_{i=1}^{N_s}$ over macro-timesteps:
\begin{align}\label{eq: discrete-signal}
	X_{k} & = X_{k-1} +   \tilde{b}\left(X_{k-1}\right)\Delta t + \sigma_x \Delta W_{k},
\end{align}
where $\tilde{b}$ is defined in~\eqref{eq: MC-averaged}. Note that \eqref{eq: discrete-signal} is the same as \eqref{eq: macro-solver} but with additive noise, which is introduced to regularize the system, for the purpose of weight calculation.

Instead of continuous observations, we assume that observations are available at discrete macro-timesteps. Write $Y^\epsilon_k \deff Y^\epsilon_{t_k}$. The discrete-time observations are
\begin{align}\label{eq: discrete-obs}
	Y^\epsilon_{k} = h\left( X^\epsilon_{k}, Z^\epsilon_{k,\left\lfloor \frac{\Delta t}{\delta t} \right\rfloor} \right) + \sigma_y V_{k}. 
\end{align} 
Observations $Y_k^\eps$ are used to update the sample by altering particle weights and resampling the system of particles. For notational consistency, we denote the particles representing the averaged $\bar X_k$  by $\bar x^i_k$.  

\begin{figure}[h]
	\centerline{\epsfig{file=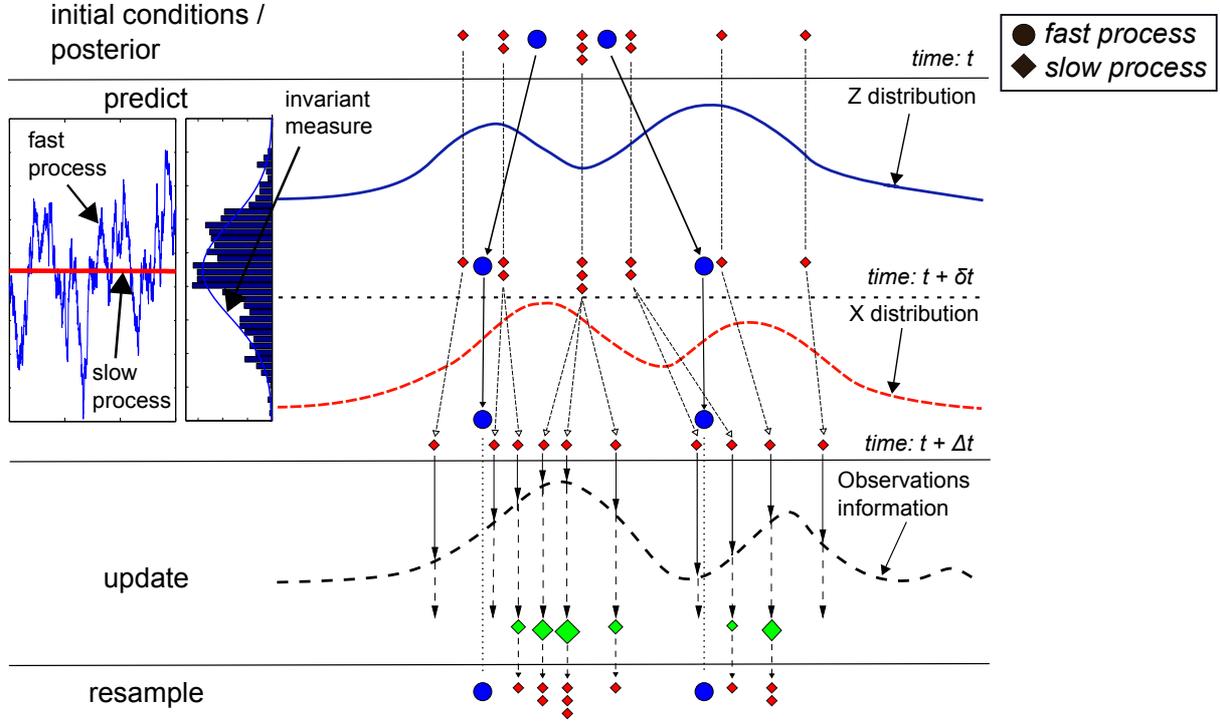,width=\textwidth}}
	\caption{HHPF illustrative scheme}
	\label{fig:hhpf_schem}
\end{figure}

The evolution of the particle system via the HHPF is given by the following steps with a graphical illustration of each step shown in Figure \ref{fig:hhpf_schem}.

\vspace{.2cm}

\begin{enumerate}
\item[(a)] {\bf initial condition}:
Samples of $N_s$ particles for the slow process, $\bar X_k$, and $N_s \times M$ particles for the fast process, $Z_{k,j}^\eps$ (subscripts $k$ and $j$ are the macro-step and micro-step indices, respectively, not to be confused with the spatial indices, superscripts $k$ and $j$, in Section \ref{S:Model}), are drawn from the initial distribution of the state variables. 
The initial distribution of the slow process, $\pi^0_{0}$, is approximated by a sample of $N_s$ particles $\{\bar{x}^i_0\}_{i=1}^{N_s}$ drawn from $\pi^0_0$, each of mass $\frac{1}{N_s}$, i.e.
\begin{align*}
\pi^0_{0} \approx \frac {1}{N_s} \sum_{i=1}^{N_s} \delta (x - \bar x_0^i).
\end{align*}
Similarly, the initial distribution of the fast process is approximated by a collection of $M$ particles $\left\{ z_0^{\eps,i,r}\right\}_{i,r=1}^{N_s,M}$ which are equally weighted ($z_0^{\eps,i,r} \in \R^n$ is the position of the $r^{th}$ particle). Note that we are only interested in the coarse-grained dynamics $\dot{\bar X}_t$ of the system. The fast process, assumed to be exponentially mixing, is spatially and temporally averaged through the implementation of the HMM. 
Thus, under the assumption of ergodicity, we can set $M<< N_s$ or even $M= 1$ without loss of accuracy in approximating the distribution of $\bar X_k$.

\item[(b)] {\bf prediction and update}: 
The prediction and update step is where HHPF differs from regular particle filters through the incorporation of dimensional reduction and homogenization techniques. 

In a regular particle filter, particles evolve independently according to the discretized signal equations which govern the behaviour over a micro-step $\delta t$. Recall that $\delta t$ is small compared to $\eps$. So, the regular filters simulate the slow process also every $\delta t$, even though it does not change much in such a small time, thus wasting the computation resources. 

The HHPF utilizes multiscale scheme of the HMM to propagate the sample forward in time. 
Particles $\left\{ \bar x_k^i \right\}_{i=1}^{N_s}$ are propagated independently over macro-timesteps $\Delta t$ using the macro-solver \eqref{eq: macro-solver}, where the averaged drift is approximated using \eqref{eq: MC-averaged}. Particle propagations in a macro-timestep interval are illustrated in the ``predict'' segment of Figure \ref{fig:hhpf_schem}. 
Within each macro-timestep interval, the particles representing the {\it fast} process, $\left\{ z_{k,j}^{\eps,i,r}: k~{\rm fixed}\right\}_{i,r=1}^{N_s,M}$, evolve according to the micro-solver \eqref{eq: micro-solver} over micro-timestep intervals $[j\delta t,(j+1)\delta t]$, $j = 1,\ldots,N_m-1$, while the {\it slow} process, $\bar X_k = x$ is fixed throughout the macro-timestep interval. Note that in the HMM scheme, $N_m<\frac{\Delta t}{\delta t}$; $N_m$ only needs to be sufficiently large for the fast process to attain an invariant measure within $[k\Delta t,(k+1)\Delta t]$. Also, the particles representing $z_{k,j}^{\eps,i,r}$ that are propagated over the micro-timesteps are in total $N_s\times M$, but $M$ can be set to be 1 by appropriately adjusting the value of $N_m$, thus the number of fast particles that need to be propagated can be $N_s$, even with implementation of the HMM scheme. The coefficient $\tilde b$ in \eqref{eq: MC-averaged} is computed by averaging over the particle locations $\left\{ z_{k,j}^{\eps,i,r}: j=1,\ldots,n_T+N_m \right\}_{i,r=1}^{N_s,M}$. 
Averaging over $z_{k,j}^{\eps,i,r}$s is performed spatially over the sample of $M$ particles and temporally over the micro-timesteps interval $[n_T\delta t,(n_T+N_m)\delta t]$.
The prediction step is given by \eqref{eq: discrete-signal}.

The particles are updated at time-steps when observations are available, as illustrated in the update segment of Figure~\ref{fig:hhpf_schem}. Only the sample of fast process $\left\{ \bar x_k^i \right\}_{i=1}^{N_s}$ is updated, since, in the multiscale scheme, we are interested only in the dynamics of the homogenized $\bar X_k$. The SIS algorithm is used for updating the sample. In propagating the particles $\{\bar{x}^i_k\}$, as described above, according to the homogenized signal dynamics over $\Delta t$, we arrive at new location of the particles $\{\bar{x}^i_{k+1}\}$. These new locations can be thought of as a sample drawn from the proposal density $q(\cdot|x_{k})$ which is same as the prior $p(\cdot|x_{k})$ of the homogenized dynamics. Particle weights are calculated sequentially according to \eqref{eq: sequential-weights} and because of our assumption of the proposal, we have
\begin{align*}
w^i_k = p\left( \left. y^\epsilon_k \right| \bar{x}^i_k\right) w^i_{k-1}.
\end{align*}
We would like to point out another difference of the HHPF compared to a regular SIS filter, that arises due to homogenization. 

Considering independent standard Gaussian noise increments for the sensor noise, i.e. $V_k \sim \mathcal{N}(0,\mathbb{I})$, in \eqref{eq: discrete-obs}, the likelihood function is Gaussian:
\begin{align}\label{eq: likelihood}
	p\left(\left. y^\epsilon_k \right| \bar{x}_k\right) \propto \exp\left\{ -\frac{1}{2}\left(y^\epsilon_k - \bar{h}\left(\bar{x}_k\right)\right)^T \left(\sigma_y \sigma_y^T\right)^{-1} \left( y^\epsilon_k - \bar{h}\left(\bar{x}_k\right)\right)\right\}.
\end{align}
Instead of the sensor function $h\left(X^\eps_k,Z^\eps_{k,\left\lfloor\frac{\Delta t}{\delta t}\right\rfloor}\right)$, the averaged sensor function
\begin{align}\label{hhpfuseIS:avgsensor}
	\bar{h}(x) = \int_{\R^n} h(x,z) \mu_x(dz)
\end{align}
is used, since we are dealing with the coarse-grained dynamics. As with the homogenized drift, the averaged sensor function is approximated by $\tilde{h}$ via the HMM. $\tilde{h}$ has the form  
\begin{equation}\label{eq: MC-averaged-h}
  \tilde h(X_k) = \frac{1}{MN_m} \sum_{r=1}^M \sum_{j = n_T}^{n_T+N_m} h(\bar{X}_k,Z^\epsilon_{k,j,r}).
\end{equation}
It is worthwhile to note that the actual available observation $Y_k^\eps$ is used instead of a fictitious averaged $\bar Y_k$ in calculating the weights. Thus the homogenized system \eqref{eq: slow-averaged-ch4} is combined with the actual observation $Y^\eps$, from which the name ``homogenized hybrid'' was derived.

After a few time steps, all the weights may tend to concentrate on a very few particles, which drastically reduces the effective sample size. This issue is addressed in the following step.

\item[(c)]{\bf resampling:} The nature of the sequential importance sampling algorithm is such that the variance of the unnormalized weights increases with each iteration (see, for example, Prop. 3, p. 7 in \citet{douc98}, Theorem, p. 285 in \citet{Kong1994}). Due to the gradual increase in weights variance over time, weights will tend to concentrate on a few particles, causing the issue of sample degeneracy. Sample degeneracy decreases the ability of the weighted sample to properly represent the target posterior distribution since only a limited number of particles will have significant weights. Additionally, it incurs the cost of propagating and storing particles with insignificant weights, which effectively do not contribute to representing the target distribution. One method of addressing this issue is to perform occasional resampling when the sample degeneracy level reaches a certain threshold. Resampling does not overcome the issue of weights degeneracy; it only serves to rejuvenate the sample by eliminating particles with insignificant weights and multiplying those with weights that significantly contribute to approximating the posterior. 

The measure of sample degeneracy can be determined by the effective sample size $N^{\mathrm{eff}}$ (see, for example, \citet{arulam02}):
\begin{align*}
	N^{\mathrm{eff}}_k \deff \frac{N_s}{1+\textrm{Var}\left(\bar{w^*}_k\right)}, \qquad \bar{w^*}^i_k \deff \frac{p\left( \left. \bar{x}^i_k \right| y^\epsilon_{0:k} \right)}{q\left( \left. \bar{x}^i_k \right| \bar{x}^i_{k-1}, y^\epsilon_{k} \right)},
\end{align*}
where $\bar{w^*}^i$ is the ``true'' weight. The exact values of $\bar{w^*}^i$ cannot be evaluated since $p( \bar{x}^i_k | y^\epsilon_{0:k} )$, the actual posterior, is not known, so the effective sample size is approximated numerically using the normalized weights by~\citet{arulam02}
\begin{align*}
	\tilde{N}^{\mathrm{eff}}_k = \frac{1}{\sum_{i=1}^M (\bar{w}^i_k)^2}.
\end{align*} 
$\tilde{N}^{\mathrm{eff}}_k \le N_s$ represents the effective sample size at timestep $k$, and a resampling procedure is carried out whenever $\tilde{N}^{\mathrm{eff}}_k$ falls below a set threshold. 

One resampling procedure is the systematic resampling procedure, as described in \citet{arulam02}, \citet{douc98}. 
The resampling procedure involves sampling with replacement on the current sample, multiplying particles with significant weights and discarding those with insignificant weights. The new sample is reinitialized with uniform weights $\frac{1}{N_s}$ for each particle. 

In addition to resampling, the issue of sample degeneracy can also be addressed by choosing a good importance sampling density $q$. As described so far, the importance density $q\left(\left. x_k \right| x_{k-1} \right)$ is based on the propagation of the sample from the previous estimation step. i.e. the observation $Y_k$ was not involved in the importance density at time $k$. In Section \ref{s:ISuseqopt} below, we propose a method of specifying an importance density $q\left(\left. x_k \right| x_{k-1}, y_k \right)$ at time $k$ that makes use of the observation at time $k$ .

\end{enumerate}

Algorithm \ref{al: hmmsis} shows a pseudo code for the HHPF. Overall, the advantages of Algorithm \ref{al: hmmsis} are: 
\begin{enumerate}
\item[(a)] The number of fast samples evaluations is greatly reduced (if the number of fast sample replicas is set as $M=1$ by choosing $N_m$ appropriately), since $N_m<\lfloor \Delta t/\delta t \rfloor$. 
\item[(b)]  The total number of timesteps is decreased due to the relatively large macro-timesteps.  
\item[(c)] The number of function evaluations are also decreased accordingly.  
\end{enumerate}

Even though there are additional function evaluations (as in equation \eqref{eq: MC-averaged}) in incorporating the HMM, these are negligible compared to the original weight calculations 
of the regular branching particle method, which have been reduced in the HHPF through \eqref{eq: MC-averaged-h}.

Although the HHPF has been adapted for multiscale computations, another issue arises in its application on complex problems with inherent chaotic nature such as that described in Section \ref{S:Model}. 
As shown in Section \ref{S:Application}, we are able to perform state estimation for a chaotic system using the HHPF as described so far, the procedure can be made more efficient by importance sampling. However, as mentioned earlier, a key aspect of the importance sampling algorithm is the choice of the sampling density $q$. In the following section, we present a method of constructing a good sampling density by introducing a forcing term in the particle evolution of \eqref{eq: discrete-signal}.

\begin{algorithm}[H]
\caption{HHPF (\citet{Park2011})}
\label{al: hmmsis}
\begin{algorithmic}
\STATE Draw samples from initial distribution: 
$\{\bar x^{i}_{k=1}\}_{i=1}^{N_s}$,$\{z^{\epsilon,i,r}_{k=1,j=1}\}_{i,r=1}^{N_s,M}$
\FOR {k = 1:number of macro-timesteps ($K$)}
\FOR {r = 1:number of replicas ($M$)}
\FOR {j = 1:number of micro-timesteps ($N_m$)}
\STATE  Solve micro-solver \eqref{eq: micro-solver}: $z^{\epsilon,i,r}_{k,j+1}$
\ENDFOR
\ENDFOR
\STATE Perform averaging \eqref{eq: MC-averaged} and \eqref{eq: MC-averaged-h}: 
$\{\tilde f(\bar x^i_k), \tilde h(\bar x^i_k)\}_{i=1}^{N_s}$
\STATE Solve macro-solver \eqref{eq: macro-solver}: 
$\{\bar x^i_{k+1}\}_{i=1}^{N_s}$
\STATE Compute weights:  
$\{\bar w_{k+1}^i\}_{i=1}^{N_s}$
\STATE Compute effective sample size: 
$\tilde{N}_{\mathrm{eff}}^{k+1}$
\STATE Resample using choice of resampling algorithm if $\tilde{N}_{\mathrm{eff}}^{k+1}<N_{\mathrm{thres}}$
\STATE Reinitialize: 
 $z^{\epsilon, i,r}_{k+1,j=1}=z^{\epsilon, i,r}_{k,N_m}$
\ENDFOR
\end{algorithmic}
\end{algorithm}

\subsubsection{HHPF using the optimal proposal}
\label{s:ISuseqopt} 

In the event that the averaged sensor function \eqref{hhpfuseIS:avgsensor} is expressible as $\bar{h}(x)=Hx+C_k$, where $C_k$ is a vector which can change with time $k$, then one can use the optimal proposal which keeps the variance of weights to minimum, as explained in \ref{ss:SIS}. Note that $\bar{h}$ is of this form if the observation function $h$ is linear and only depends on the slow components. Only the following changes need to be made to the algorithm:
\begin{itemize}
\item Instead of propagating the particles $\bar x^i_{k+1}$ by using \eqref{eq: discrete-signal}, we propagate according to
\begin{align}\label{ISforHHPF}
	X_{k} & = X_{k-1} +   \tilde{b}\left(X_{k-1}\right)\Delta t + \alpha(X_k,y_k-C_k)+ \hat{\sigma}_x \Delta W_{k},
\end{align}
where $\hat{\sigma}_x$ is such that $\hat{Q}\deff(\hat{\sigma}_x\hat{\sigma}_x^T)=((\sigma_x\sigma_x^T)^{-1}+H^T(\sigma_y\sigma_y^T)^{-1}H)^{-1}$, and 
$\alpha(X_k,y_k)=\hat{Q}H^T R^{-1}(y_k-Hf(X_{k-1}))$, where $f(x)=(x +   \tilde{b}\left(x\right)\Delta t)$.
\item The weights are updated according to
\begin{align}
w^i_k\,\,&\propto\,\, w^i_{k-1}\exp\left\{-\frac12 (y_k-C_k-Hf(x_{k-1}^i))^T\hat{R}^{-1}(y_k-C_k-Hf(x_{k-1}^i))\right\}, \nonumber \\
\hat{R}^{-1}&=R^{-1}\left(\mathbf{1}-H\hat{Q}H^TR^{-1}\right).
\end{align}
\end{itemize}

\section{Application}
\label{S:Application}
Based on the results of homogenization and optimal importance sampling, we have developed a new lower-dimensional particle filter, the HHPF, for state estimation in nonlinear multi-scale systems. In this section, we illustrate the HHPF's potential for state estimation in a high-dimensional complex problem by applying the HHPF algorithm to the Lorenz '96~(\citet{lorenz96}) atmospheric model (see Section \ref{S:Model}) to estimate the {\it slow} variables. The HHPF algorithm is implemented in discrete time, using SIS, as presented in Section \ref{s:HHPF}. The model parameters for application of the HHPF are as follows:
$K = 36$ $J = 10$, $F_x = 10$, $h_x = -0.8$, $h_z = 1$, $\epsilon = 1/128$.

The model was first simulated for $40960$ timesteps of size $2^{-11}$ starting from arbitrary initial conditions to represent the ``true'' signal $X^\epsilon$ that was to be estimated. The timestep value of $2^{-11}$ was picked such that it is small enough compared to the separation parameter $\epsilon$ to ensure numerical stability. The Lorenz '96 model given by \eqref{eq: lorenz-slow}, \eqref{eq: lorenz-fast-noise} was integrated using a split timestepping scheme: fourth order Runge-Kutta scheme for the deterministic drift and the Euler-Maruyama scheme for the stochastic parts. These schemes were selected for simplicity of implementation in the numerical experiments, but of course, coarser timesteps may be used by implementing higher order integration schemes. Observations generated from the true states were of the form assumed in Section \ref{s:ISuseqopt}: $Y^\epsilon_t = HX^\epsilon_t + B_t$, i.e. depends linearly on $X^\epsilon$, perturbed by a standard Gaussian noise. Observations were generated every $128$ timesteps, i.e. at every timestep of size $2^{-4}$. This is the size of the macrotimestep $\Delta t$ that we select for the HMM integration scheme. This means that we are considering the case where observations are available sequentially (at every timestep) at the timescale that we had chosen for the numerical integration of the slow process. In~\citet{lorenz96}, the timestep chosen for numerical integration of the multiscale system with $\epsilon = 0.01$ was $0.05$, which corresponds to $36$ minutes in real time. We follow~\citet{fatkullin04} in selecting $\Delta t = 2^{-4}$ and $\delta t = 2^{-11}$ for the HMM scheme for $\epsilon = 2^{-7}$. Based on~\citet{lorenz96}'s scale, these timesteps approximately correspond to $45$ minutes and $35$ seconds, respectively, in real time and the duration of the model simulation and data assimilation experiments correspond to 10 days (14400 minutes). On the real time scale, the assumption of sequentially available observation data corresponds to observation data collection every $45$ minutes, which is not an unrealistic assumption. 

We consider two cases for the linear observation matrix $H$. The first is $H=I_{K\times K}$, where $I$ is the identity matrix, i.e. all slow states are observed (non-sparse observations). The second is $H_{ij}=1$ if $i=j$ and $i$ odd, and otherwise $H_{ij}=0$ , i.e. only the odd-indexed slow dimensions are observed (sparse observations). Observation noise covariance matrix is assumed to be the identity matrix, i.e. observation noise in each dimension is independent of the rest. We chose the slow and fast signal noise covariance matrices to be $1$ on the diagonals and $0.5$ on the sub- and super-diagonals as in~\citet{Leeuwen2010}, with the fast noise scaled by $\epsilon^{-1/2}$. Initial conditions for the true signal are arbitrarily chosen from mean zero normal distributions with variances $3$ and $5$ for the slow and fast processes, respectively. The filtering objective is to estimate the {\it slow} states using sequentially available observations. We discuss the numerical experiments in the following. The numerical experiments were performed using MATLAB v. 7.11.0.584 (R2010b), without explicitly employing parallel processing capabilities, on an Intel Xeon DP Hexacore X5675 3.07 GHz processor with 12$\times$4 Gb RAM. 

\subsubsection{Optimized HHPF with linear observations}

In the first set of numerical experiments, we consider the case of non-sparse observations and consider two variants of the HHPF to study the effects of the optimal particle propagation for linear observation functions discussed in Section \ref{s:ISuseqopt}. The SIS algorithm is used in both variants of the HHPF. In one variant, the proposal density used for the importance sampling procedure is generated using particles propagated using the optimal drift and diffusion as given in \eqref{SIS:example_evolve_particles}, with weights updated accordingly. In the other variant, proposal density is the prior density generated by propagating particles directly according to \eqref{eq: discrete-signal}. We call the first one the \emph{optimized HHPF} and the second one the \emph{direct HHPF}. For the HMM scheme, the HMM window was set at $m_N = 64$ microtimesteps, and number of microtimesteps skipped to ignore transient effects is $m_T = 32$. The number of replicas is set at $m_r = 1$. 

\begin{figure}[h]
	\centerline{\epsfig{file=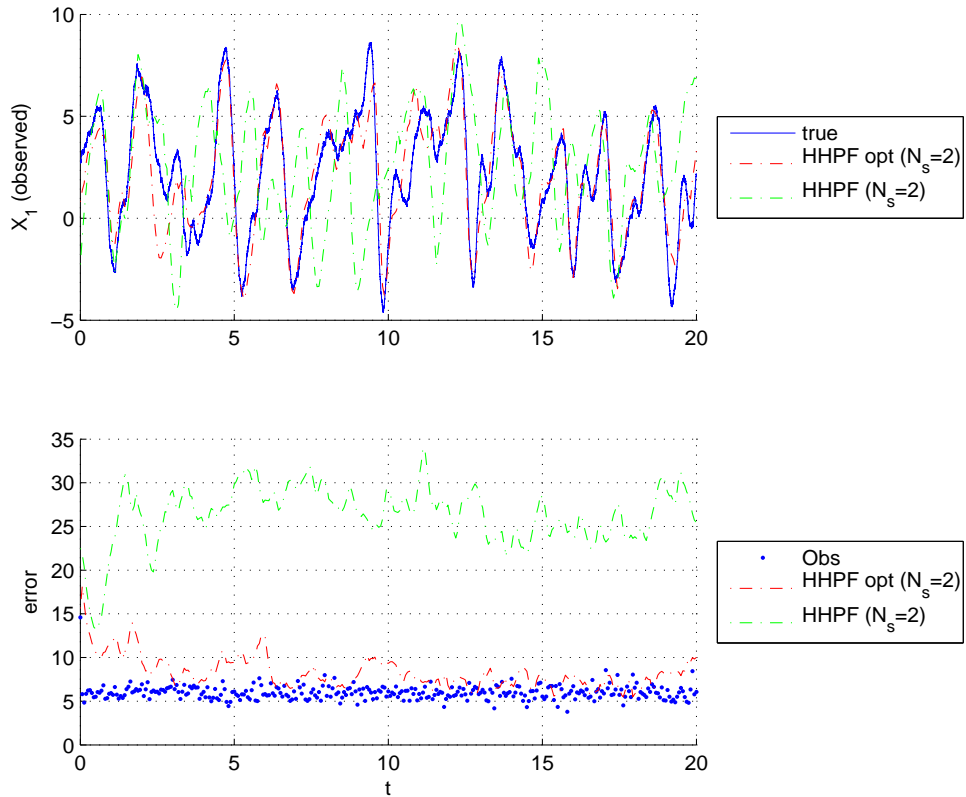,width=.8\textwidth}}
	\caption{Non-sparse observations, optimized and direct HHPFs, $N_s=2$. The optimized HHPF estimated the truth fairly well but was at best as good as observations. Increasing $N_s$ led to estimates that were better than obervation data.}
	\label{f: HI-compare-HHPFs-2part}
\end{figure}

\begin{figure}[h]
	\centerline{\epsfig{file=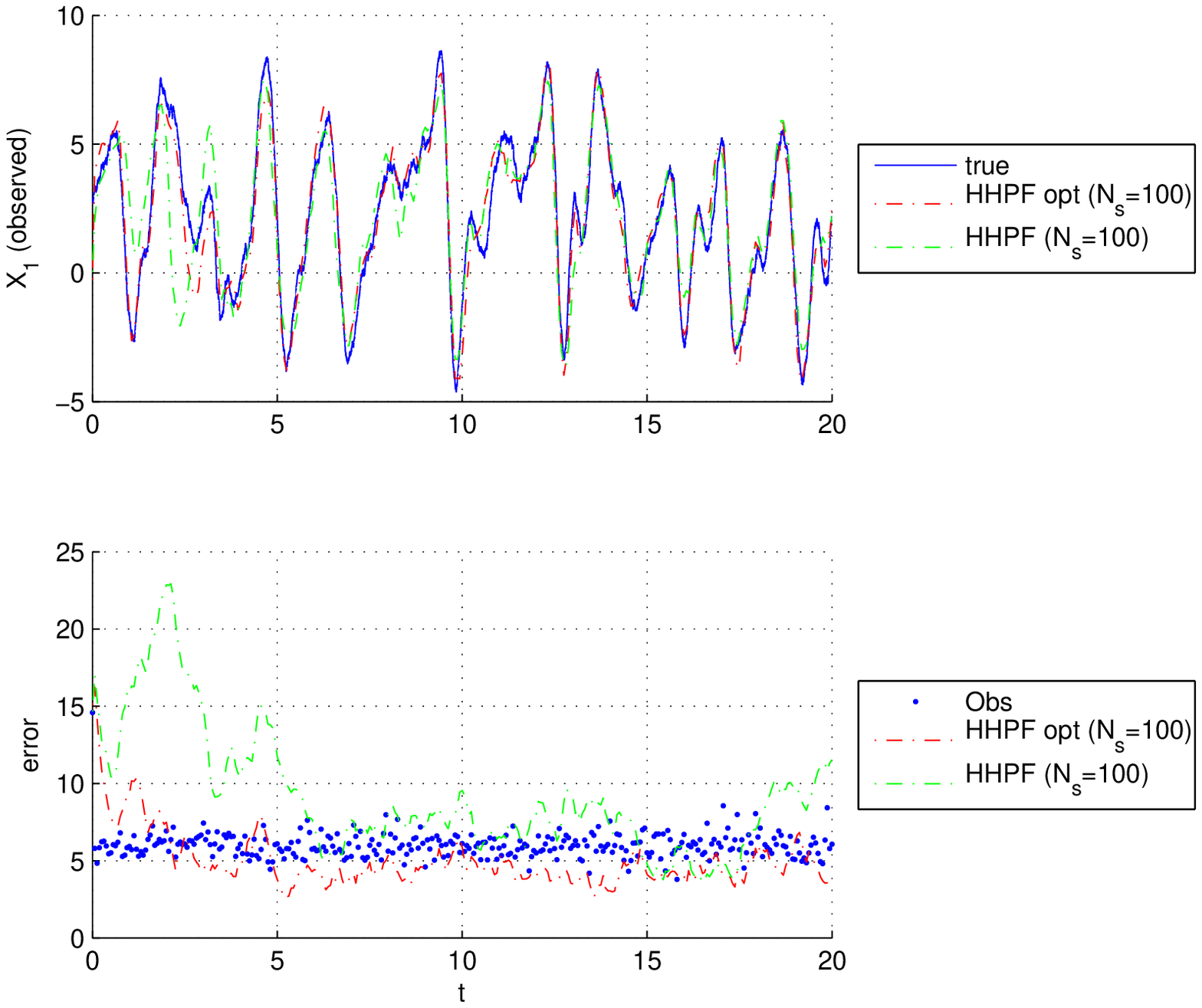,width=.8\textwidth}}
	\caption{Non-sparse observations, optimized and direct HHPFs, $N_s=100$. The optimized HHPF estimate was better than observation data but the direct HHPF was still not as good. The direct HHPF was able to match the optmized one with $N_s=600$ at the cost of longer computation time.}
	\label{f: HI-compare-HHPFs-100part}
\end{figure}

Numerical experiments were performed using both HHPFs with varying sample sizes starting from $N_s = 2$ to $N_s = 400$. Both filters achieved better accuracy with increasing $N_s$, as they should, and Figures \ref{f: HI-compare-HHPFs-2part} and \ref{f: HI-compare-HHPFs-100part} each show the comparison of one dimension of the true state with its estimates using the HHPFs and the corresponding estimation errors. In both upper figures, blue curve represents the true state, broken red curve represents the estimate using the optimized HHPF, and green curve represents that of the direct HHPF. In the lower figures, blue curve represents the observation error, broken red curve and blue curves represent the estimation error of the optimized and direct HHPFs, respectively. The error shown is the absolute error over all $36$ slow dimensions, i.e. ${{\rm error}}_t = \sqrt{\sum_{k=1}^{36} (\hat{X}^{k}_t-X^{k, \epsilon}_t)^2}$. 
Figure \ref{f: HI-compare-HHPFs-2part} shows the HHPFs using \emph{just $2$ particles} and the optimized HHPF was seen to perform fairly well in estimating the true state, and the error plot showed that the estimate was only slightly worse than the observation error. This is not surprising, since all the slow states were observed and the particles were driven towards those states as indicated by the observations. Estimation error was contributed to in part by the stochasticity in the system and the observation noise, as well as the error due to approximation by numerical homogenization. The direct HHPF however fails to estimate the truth with $N_s=2$. The estimate initially converged to the truth but diverges at around $t=2.5$ and fails to recapture it after that. Both the optimized and direct HHPFs took just $5$ seconds to run over the entire interval of $320$ macrotimesteps, which is a good performance for the optimized HHPF. But considering that the estimates were at best only as good as the noisy observations, the most computationally efficient choice would be to just use observations directly if observations of all states are available. 

By increasing the number of particles used, we see in the error plot of Figure \ref{f: HI-compare-HHPFs-100part} that the optimized HHPF was able to provide a better estimate of the truth than what was known directly from the observations. The run time for the experiment presented in the figure was $135$ seconds, which is the typical time recorded for the optimized HHPF with $N_s=100$. The estimate from the direct HHPF improved but was at best as good as the observation with $N_s=100$. Increasing $N_s$ up to $400$ improved both filters' estimates but not significantly for the optimized HHPF. The direct HHPF was seen to be able to match the optimized HHPF with $N_s=100$ by using $600$ particles but the run time required was $639$ seconds. The second and third columns of Table \ref{t: HI-runtimes} show the run times for different sample sizes for a typical experiment using the HHPFs. For each fixed $N_s$, run times for the optimized and direct HHPFs were observed to be within the same range, which was expected, since the only additional function evaluations in the optimized version is the calculation of the ``nudging'' correction that is just a linear combination of observation vector with particle locations. Occasional drastic variation in run times between the optimized and direct HHPFs, for example for $N_s=100$ and $N_s=400$, can be explained based on the number of times the resampling procedure were required to be performed in each filter. In each case, resampling may have been required more frequently due to the initial sample being significantly far from the truth, hence more resampling had to be performed in the beginning of the run, or at some point in time, a large number of particles were forced significantly far from the true state by the stochastic forcing. The second scenario is less likely to happen since the signal noise amplitude is small in comparison with that of the states. Resampling, however, will not drastically increase for the optimized HHPF because particles are simultanesously driven towards the truth based on the observations through the optimal propagation procedure.
   
\begin{figure}[h]
	\centerline{\epsfig{file=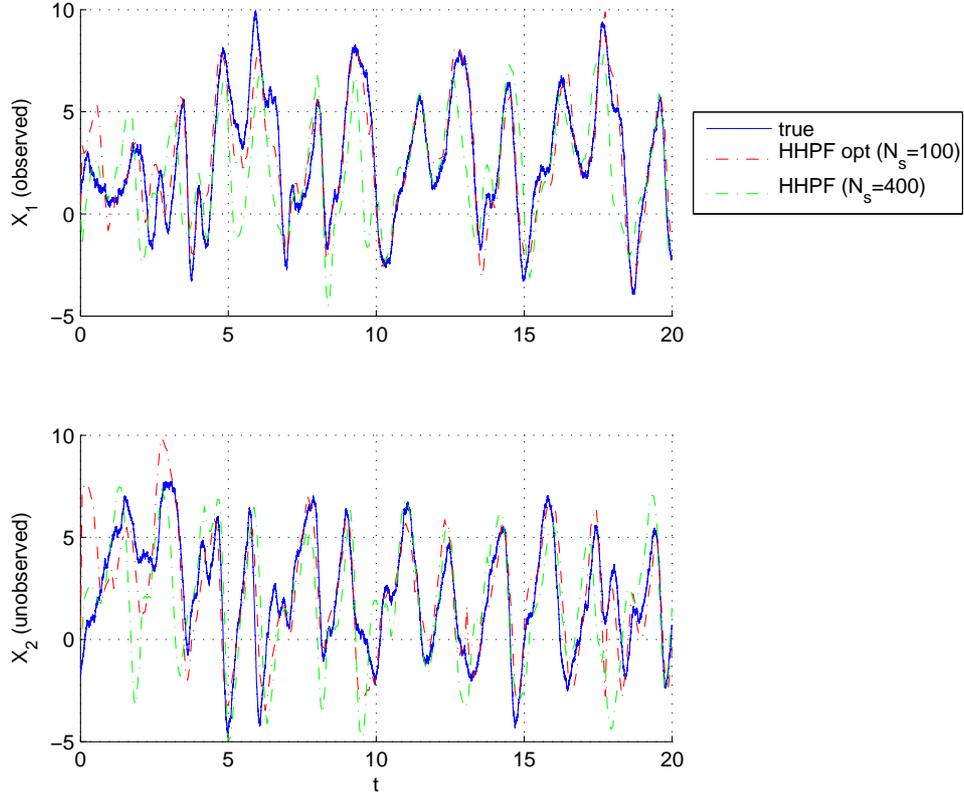,width=.8\textwidth}}
	\caption{Sparse observations, optimized ($N_s=100$) and direct ($N_s=400$) HHPFs. The upper plot shows the estimates of an observed state, the lower plot shows that of an unobserved state. The unobserved state was estimated well by the optimzied HHPF with $N_s=100$. Even with $N_s=400$, The direct HHPF captures the fluctuations in the truth but did not follow the trajectory well.}
	\label{f: Hodd-compare-HHPFs-100part-400part}
\end{figure}

\begin{figure}[h]
	\centerline{\epsfig{file=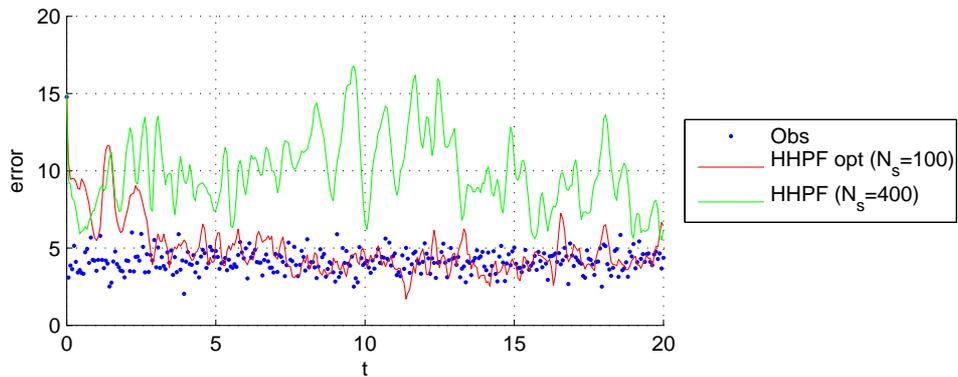,width=.8\textwidth}}
	\caption{Sparse observations, estimation error of the observed states for the optimized ($N_s=100$) and direct ($N_s=400$) HHPFs compared with observation error. The optimized HHPF estimates were as good as observations (but the optimized HHPF also provided estimates of the unobserved states so more information is gained by using the filter instead of just observation).}
	\label{f: Hodd-compare-HHPFs-100part-400part-errors}
\end{figure}

The same numerical experiments were also performed for the case of sparse observations (i.e. observing only the dimensions with odd index), to study the performance of the optimized HHPF in estimating hidden states. The same trend as for the case of non-sparse observations was observed in comparing the estimates using the optimized and direct HHPFs. For fixed $N_s$, the optimized HHPF achieved better accuracy than the direct version, although both estimates displayed higher estimation errors compared to the non-sparse observation case at the same $N_s$ due to the presence of hidden states. For this case, the optimized HHPF estimate was able to match the observation data of the observed states using $20$ particles, and estimate the unobserved states as well. Figure \ref{f: Hodd-compare-HHPFs-100part-400part} shows the comparisons of the estimates from the HHPFs, using $100$ particles for the optimized and $400$ particles for the direct,  with the truth. The upper plot shows the estimates of an observed state and the lower shows those of an unobserved state. The optimized HHPF (broken red curve) provided good estimates of all the states, including the unobserved ones and the corresponding error plot in Figure \ref{f: Hodd-compare-HHPFs-100part-400part-errors} shows that the estimation was as good as the observation (of course, observations only observed half the states and the error shown only compares error in the observed dimensions;  the HHPF estimates included the unobserved dimensions, shown to capture the truth in the lower plot in Figure \ref{f: Hodd-compare-HHPFs-100part-400part}). The direct HHPF, was also able to capture the shape of the true fluctuations in the observed and unobserved dimensions, but in the observed dimensions, the estimates were poorer than the observations, even with $400$ particles. Increasing sample size further up to $N_s=800$ did not lead to significant improvement. The run times and the trend of increase in run time with sample size for the sparse observations experiments were the same as that of the non-sparse observations experiment shown in Table \ref{t: HI-runtimes}.

The homogenized filters comparison experiments showed that the optimized HHPF displayed significant estimation performance over the direct version for a fixed sample size. This indirectly led to improvement in computation time in comparison with the direct HHPF, in the sense that for the same or even better level of estimation error, the optimized HHPF could be implemented using smaller sample size than the direct version, hense reducing computational costs. In the discussion here and in Table \ref{t: HI-runtimes}, we have only considered the comparisons of the HHPFs with $N_s$ up to $400$, beyond which the run time advantage of the HHPF over other unhomogenized nonlinear filters, for the same level of accuracy, is lost. In the next set of experiments, we compare the optimized HHPF with two other nonlinear filters.

\subsubsection{Comparison of the optimized HHPF with other nonlinear filtering schemes}
\label{subssubs: nonlinear-filters-comparison}

In the second set of numerical experiments, we compared the optimized HHPF with two other nonlinear filtering schemes: the ensemble Kalman filter (enKF) and a particle filter without homogenization. The aim was to compare the run time reduction in the HHPF, due to the implementation of a homogenization scheme, relative to unhomogenized nonlinear filters and assess the trade-off in estimation accuracy due to homogenization. We do not expect the optimized HHPF to outperform an unhomogenized filter in terms of estimation accuracy, but the optimized HHPF should possess comparable estimation capabilities to unhomogenized filters, with the advantage of the HHPF being that it requires shorter run times. 

Since observations were generated at macrotimesteps of size $2^{-4}$, we have observations being sparse in time as well for the unhomogenized filters, which use the same timestep size of $2^{-11}$ as the model simulation. The enKF \citet{Evensen1996} was implemented directly without modifications, by propagating the ensemble forward in time according to the model dynamics and performing information updates at timesteps when observations are available. For the particle filter, we implement a modified SIS particle filter algorithm developed by~\citet{Leeuwen2010} that is designed to accommodate for observations that are sparse in time. This particle filter is similar to the optimized HHPF presented here, in the sense that it uses the presently available observation to construct a better proposal density at a present time by driving particles in between observation timesteps using a time exponential function that is proportional to the model noise covariance and the distance of the intermediate particle locations from the observed state. For details and better insight to this particle filter, see~\citet{Leeuwen2010}, and from here on, we will denote this particle filter as just PF, but it is implied that it is the particle filter adapted for sparse-in-time observations.

\begin{figure}[h]
	\centerline{\epsfig{file=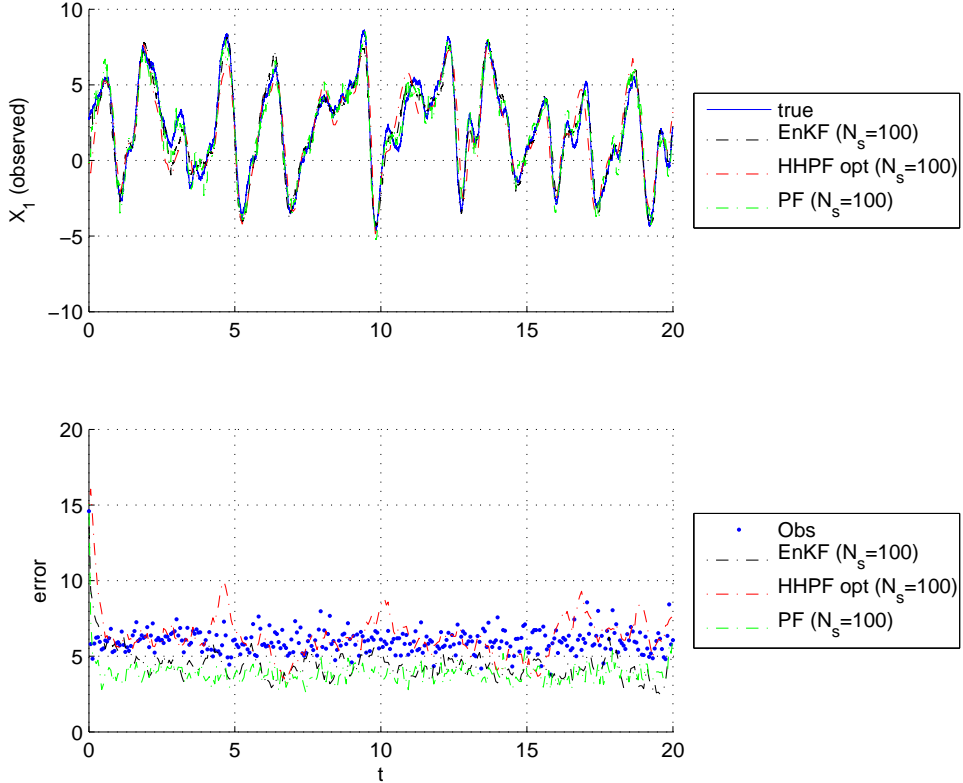,width=.8\textwidth}}
	\caption{Non-sparse observations, EnKF, PF and optimized HHPF comparison at fixed $N_s=20$. The estimate of the optimized HHPF is comparable to those of the EnKF and the PF, but is obtained in a shorter run time.}
	\label{f: HI-compare-filters-20part}
\end{figure}

\begin{figure}[h]
	\centerline{\epsfig{file=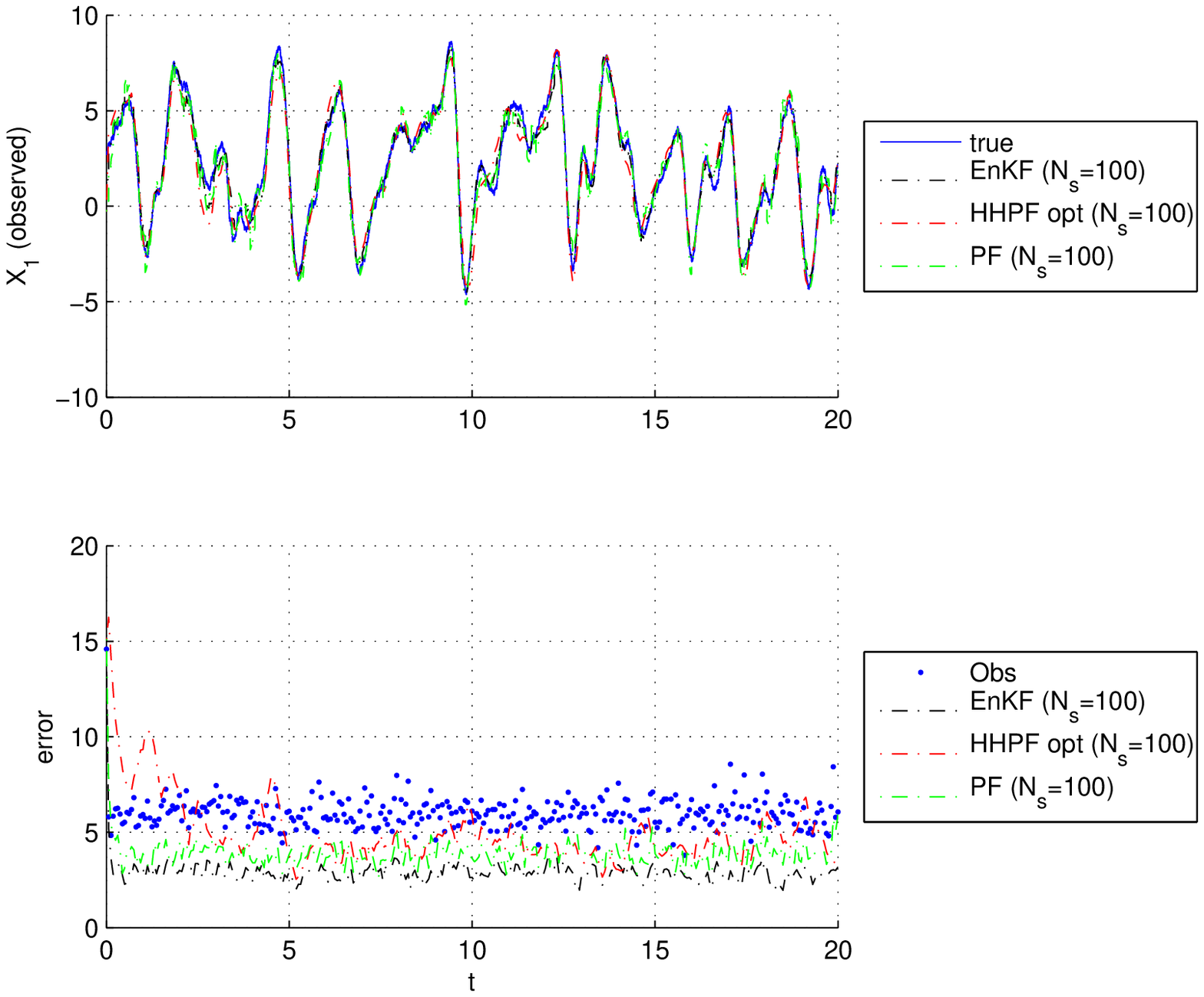,width=.8\textwidth}}
	\caption{Non-sparse observations, estimation error comparison the EnKF, PF and optimized HHPF comaprison at fixed $N_s=100$. The optimized HHPF could match the PF, withthe EnKF being slightly better, but the optimized HHPF required the least computation time.}
	\label{f: HI-compare-filters-100part}
\end{figure}

Similar numerical experiments as for the comparison of the optimized and direct HHPFs were performed for the enKF and the PF and the estimation results were compared with those of the HHPFs. We will first discuss the case of non-sparse (spatially) observations, i.e. the case $H=I_{K\times K}$. The PF was able to provide good estimates of the truth with a sample size of just $2$. Estimation error decreased as $N_s$ was increased to $20$ and was not seen to further decrease significantly as $N_s$ was increased from $20$. Figure \ref{f: HI-compare-filters-20part} shows the estimates of one dimension of the true state using the EnKF, the optimized HHPF, and the PF with $N_s=20$, and their corresponding estimation errors. The blue curve is for the truth, the black for the EnKF, red for the optimized HHPF, and green for the PF. The optimized HHPF displayed the highest estimation error, its estimate being as good as the observation. The EnKF and PF estimates are better than the observation, with estimation errors of equal magnitude. However, when considering the run times, the optimized HHPF took 30 seconds while the EnKF and PF took $540$ and $1757$ seconds, respectively. Even with $N_s=2$, the PF took $1727$ seconds due to the timestep size and the functional evaluations required for particle weight calculations, as well as the resampling procedures. Additionally, based on the error plot in Figure \ref{f: HI-compare-filters-20part}, the estimation error of the optimized HHPF was not much worse than those of the EnKF and PF, and the estimate trajectory followed the truth very well apart from slight over- and under-shoots at local maxima and minima. This indicates that the optimized HHPF required less run time compared to the unhomogenized filters at a comparable level of estimation error. Increasing sample size showed that the optimized HHPF was almost as good as the PF at $N_s=100$, as shown in Figure \ref{f: HI-compare-filters-100part}, with the EnKF being slightly better than both particle filters. The key point is that using $100$ particles, the optimized HHPF took $134$ seconds while the EnKF and PF took $2169$ and $920$ seconds, respectively. Considering that the levels of estimation errors are almost equal, the optimized HHPF provided significant advantage in terms of computation time. Further increasing the sample size to $400$ enabled the optimized HHPF to match the PF. Beyond $N_s=400$ however, the optimized HHPF lost its computation time advantage, for the EnKF could be implemented using $N_s=50$ or $100$ with about the same level of estimation accuracy and computation time. 

\begin{figure}[h]
	\centerline{\epsfig{file=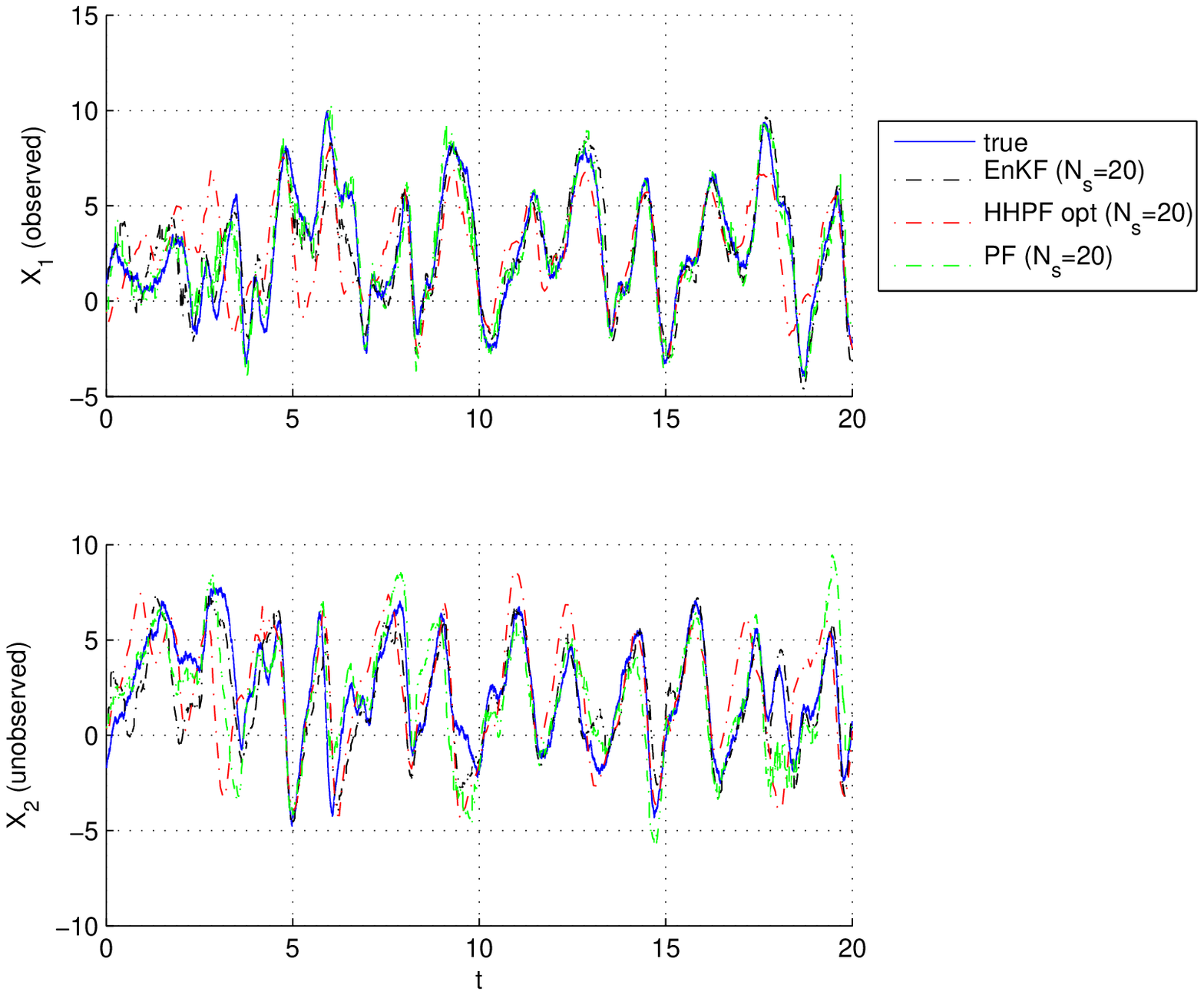,width=.8\textwidth}}
	\caption{Sparse observations, EnKF, PF and optimized HHPF comparison at fixed $N_s=20$. The optimized HHPF did not perform as well as the EnKF or PF.}
	\label{f: Hodd-compare-filters-20part}
\end{figure}

\begin{figure}[h]
	\centerline{\epsfig{file=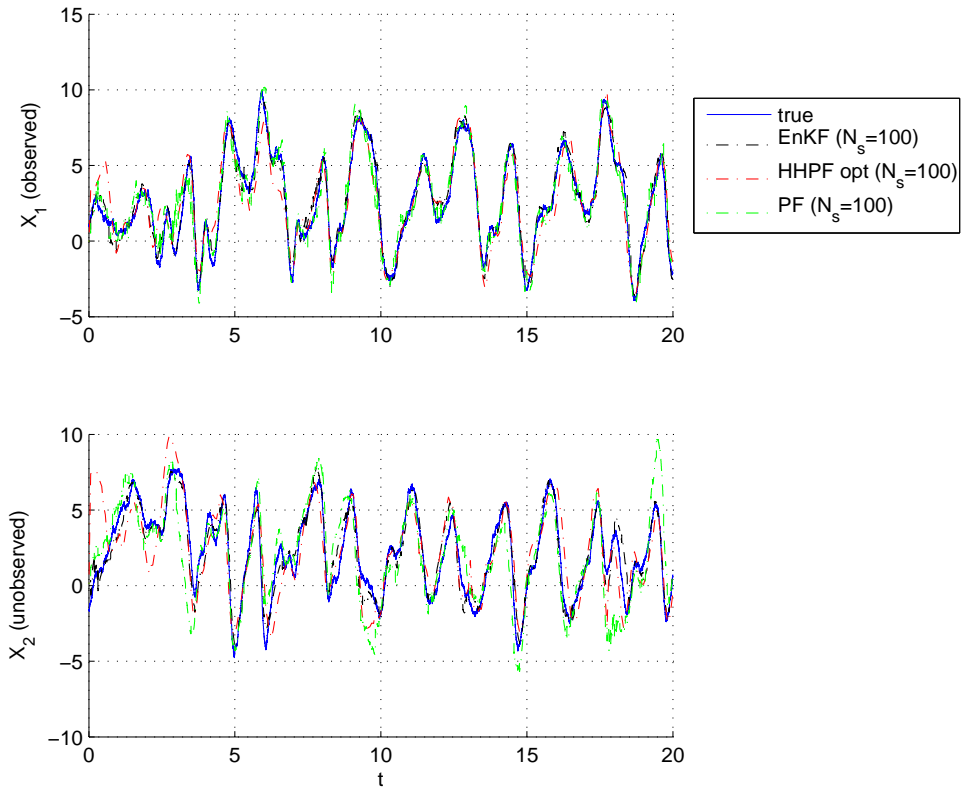,width=.8\textwidth}}
	\caption{Sparse observations, EnKF, PF and optimized HHPF comparison at fixed $N_s=100$. The optimized HHPF performed as well as the PF, but in shorter time than both the EnKF and PF.}
	\label{f: Hodd-compare-filters-100part}
\end{figure}

\begin{figure}[h]
	\centerline{\epsfig{file=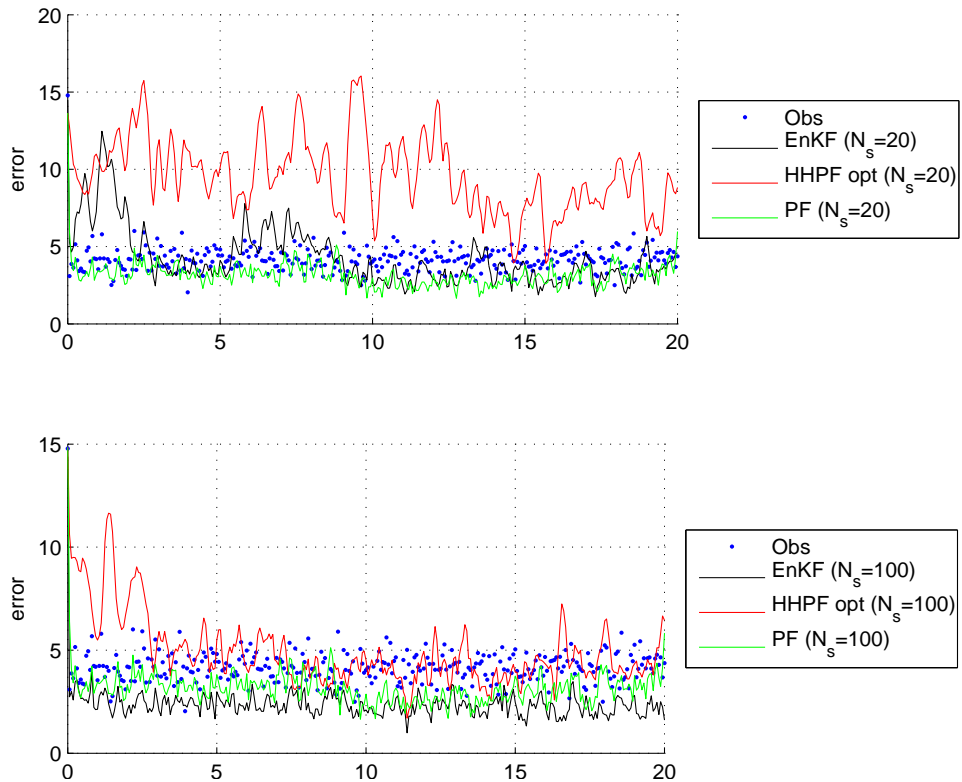,width=.8\textwidth}}
	\caption{Sparse observations, comparison of estimation errors of the EnKF, PF and optimized HHPF with observation error.}
	\label{f: Hodd-compare-filter-errors-observed}
\end{figure}

\begin{figure}[h]
	\centerline{\epsfig{file=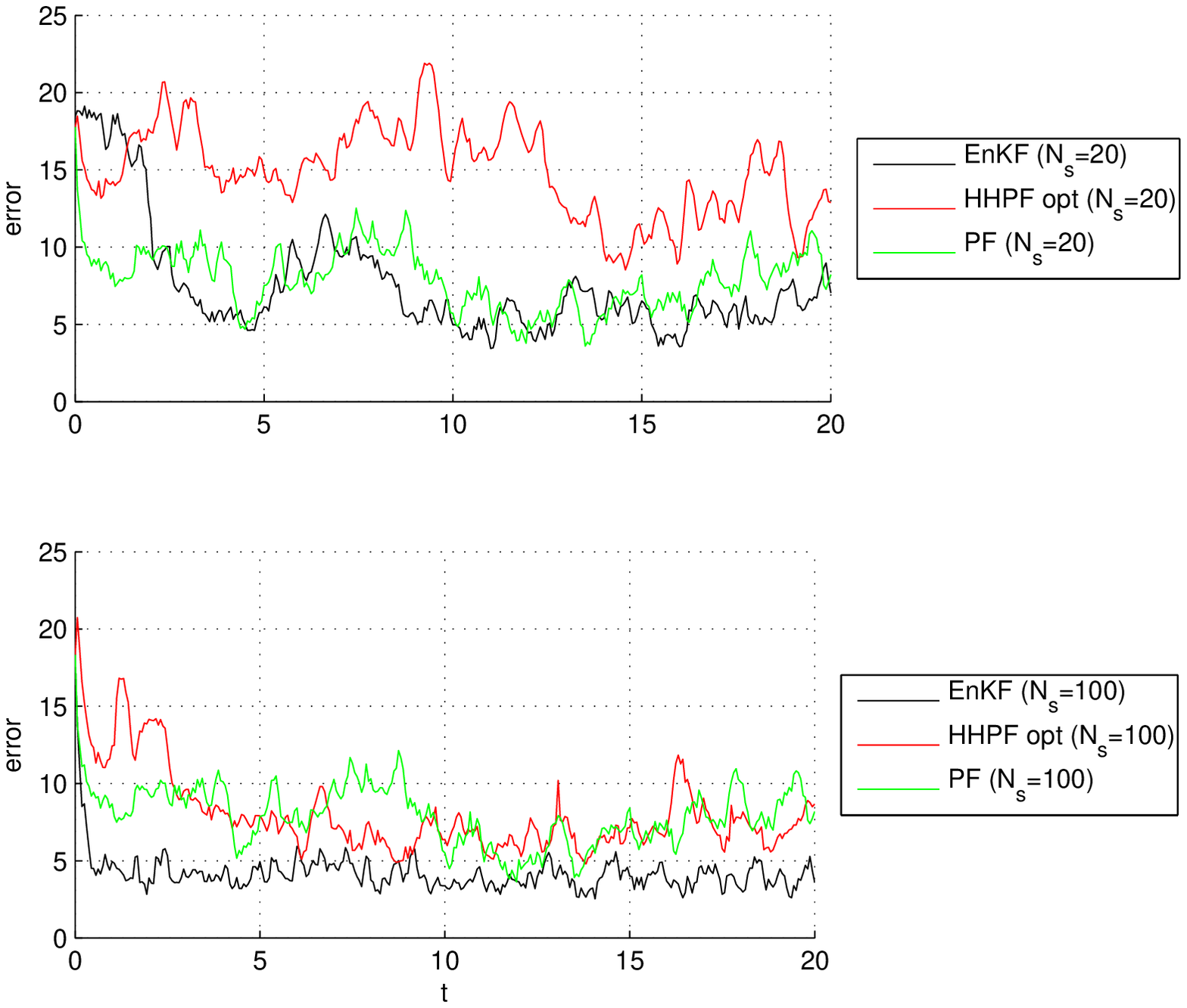,width=.8\textwidth}}
	\caption{Sparse observations, comparison of estimation errors of the EnKF, PF and optimized HHPF with observation error. Estimation error of the optimized HHPF and the PF are of the same magnitude at $N_s=100$. The optimized HHPF still required less computation time than the EnKF and PF even with the EnKF and PF being implemented using minimum $N_s$ possible.}
	\label{f: Hodd-compare-filter-errors}
\end{figure}

Similar numerical experiments were again performed for the case of sparse observations (i.e. observing dimensions with odd index) and Figures \ref{f: Hodd-compare-filters-20part} and \ref{f: Hodd-compare-filters-100part} show the comparisons of the estimates of the observed and unobserved states from the filters using $20$ and $100$ particles, respectively. With $20$ particles, the EnKF and PF provided slightly better estimates of the observed states than the observation, but the optimized HHPF performed rather poorly in comparison. However, at $N_s=100$, the optimized HHPF's performance improved significantly, performing as well as the PF. Figure \ref{f: Hodd-compare-filter-errors-observed} shows the estimation error over the observed dimensions and, to compare the filter estimate over all slow dimensions, Figure \ref{f: Hodd-compare-filter-errors} shows the estimation error over all slow dimensions for each filter. The PF estimate converged to the truth faster but the optimized HHPF estimate eventually became as good as the PF's, with the EnKF's being the best. The key point is again that the optimized HHPF took $134$ seconds while the EnKF and PF took $920$ and $2169$ seconds respectively. Even if we considered using the EnKF with $N_s=20$ ($540$ seconds) and the PF with $N_s=2$ ($1727$ seconds), both of which provided relatively low-error estimates, the optimized HHPF with $N_s=100$ is still faster. So, even in the sparse observations case, the optimized HHPF could be implemented in shorter time with estimation error comparable or equal to the EnKF and the PF.  

We do not claim that the HHPF is better than the EnKF or the PF; as shown, the unhomogenized filters provided lower estimation error than the homogenized filter at low fixed $N_s$. As mentioned in the discussion of the previous set of experiments, the accuracy of the optimzied HHPF could be increased by increasing $N_s$, but beyond $N_s=400$, it loses computation time advantage over the EnKF. However, in terms of computation time and cost of storage, the optimized HHPF displayed advantage over the unhomogenized filters, with comparable level of accuracy. The EnKF would have been the better choice of filter if the lowest possible estimation error was required over computational time. Additionally, the PF was designed to accommodate temporally sparse observations, as it has been shown to do in \citet{Leeuwen2010} and here. This capability still needs to be incorporated in the HHPF as presented here, because in most real time applications, the time interval for availability of observation data can be greater than the $45$ minute-interval assumed here.    

In comparing the computational times of the numerical experiments using the EnKF, the PF and the optimized HHPF so far, we have not taken into account the cost of computing the homogenized observation function $\bar{H}$. In addition to $H$ being a constant matrix, we also assumed that the sensor observed only the slow states, hence the observation function was independent of the fast process. So, the homogenized observation function is the same as the unhomogenized one. However, we do not expect the cost of evaluating the homogenized observation function to drastically affect the computational time advantage of the optimized HHPF over other unhomogenized nonlinear filters.  

\begin{center}
\begin{table}[h]
	\begin{tabular}{| >{\centering\arraybackslash}m{1in}| >{\centering\arraybackslash}m{1in}| >{\centering\arraybackslash}m{1in}| >{\centering\arraybackslash}m{1in}| >{\centering\arraybackslash}m{1in}|}
	\hline 
	$N_s$ & Opt. HHPF & Direct HHPF & PF & EnKF \\
	\hline 	
	2 & 5 & 5 & 1727 & N/A \\
	10 & 16 & 15 & 1716 & N/A \\ 
	20 & 30 & 20 & \pscirclebox[linecolor=red]{1757} & \pscirclebox[linecolor=red]{540} \\
	50 & 67 & 46 & 1936 & 651 \\
	100 & \pscirclebox[linecolor=red]{134} & 86 & 2169 & 920 \\
	200 & 177 & 176 & 3002 & 1415 \\
	400 & 401 & 539 & 3591 & 3152 \\
	\hline
\end{tabular}
\caption{Typical computation times (in sec.) for different sample sizes for the different nonlinear filters, in the case of non-sparse observations. We see that for a fixed $N_s$, the HHPF required less computational time. Circled are the computation times corresponding to the sample sizes that led to the same levels of estimation accuracy for the three different filters compared in Section \ref{subssubs: nonlinear-filters-comparison}. For the same level of accuracy, the optimized HHPF required a larger sample size but still performed in less time. Times for the case of sparse observations are of the same magnitudes and display the same trend.}
\label{t: HI-runtimes}
\end{table}
\end{center}

\section{Conclusion and future directions}
\label{S:Conclusion}

\subsubsection{Future directions}
In chaotic systems, such as the one studied in this paper, the transients become irrelevant from the dynamical systems point of view and the 
motion of the solution settles typically near a subset of the state space, called an attractor.
However, in the data assimilation problem that is of interest in this paper, we are interested in the transients and, in particular, in directions that are stretched by the transient dynamics. This sensitivity to initial conditions are characterized by the finite time Lyapunov exponents, which are determined by the behavior of two neighboring orbits or the two point motion of the nonlinear systems. 

We are mostly interested in filtering \emph{deterministic} chaotic systems, and the particle filtering methods developed above won't work without
the addition of noise, because knowing the initial conditions $X_0$, the distribution of $X_t$ is a Dirac measure. Therefore we add Gaussian noise \emph{artificially}. 
Hence, the  finite time Lyapunov exponents may be handy in deciding how we choose the magnitude of the noise. 

The second question then arises as to the property of the sensor function, in the linear case, the span of the observation matrix. 
As pointed out by Lorenz~\citet{Lorenz1998} and~\citet{Palmer98}, who have coined the word  ``adaptive'' or
``targeted'' observations, the sensors should be deployed at any given time, if the
data that they gather are to be most effective in improving the analysis and forecasts.

The results presented in this paper assume that the observations are available every 45 minutes. The extension of this work that deals
with every combination of the spatial and temporal sparsities, performing intermittent in time sparse data assimilation that mimics a 
global weather model is presented in~\citet{LNPY}. The particle method presented in~\citet{LNPY} consists of control terms in the 
``prognostic" equations, that nudge the particles toward the observations, specially in the sparsest situation of $\frac{K}{4}$ observations in every 48 hours,
and shows that control methods can be used as a basic and flexible tool for the construction of the proposal density inherent in particle filtering.


\vspace{20pt}
\textbf{Acknowledgement.}  Nishanth Lingala, N. Sri Namachchivaya, and Hoong C. Yeong are supported by the National Science Foundation under grant number EFRI 10-24772 and by AFOSR under
grant number FA9550-08-1-0206. Nicolas Perkowski is supported by a Ph.D. scholarship of the Berlin Mathematical School. Part of this research was carried out while Nicolas Perkowski was visiting the Department of Aerospace Engineering of University of Illinois at Urbana-Champaign. He is grateful for the hospitality at UIUC. The visit of Nicolas Perkowski was funded by NSF grant number EFRI 10-24772 and by the Berlin Mathematical School. Any opinions, findings, and 
conclusions or recommendations expressed in this paper are those of the authors and do not necessarily reflect the views of the National Science Foundation.

\begin{footnotesize}
\bibliography{bib-long}
\bibliographystyle{abbrvnat}
\end{footnotesize}

\end{document}